\documentclass[aps,prb,twocolumn,showpacs,preprintnumbers,superscriptaddress,amsmath,amssymb,longbibliography]{revtex4-2}

\usepackage{graphicx}
\usepackage{dcolumn}
\usepackage{bm}
\usepackage{hyperref}
\usepackage[version=4]{mhchem}
\usepackage{amssymb}
\usepackage[svgnames]{xcolor}
\usepackage{amsmath}
\usepackage{float}
\usepackage{caption}
\usepackage{subcaption} 
\usepackage{overpic}
\usepackage{xurl}
\usepackage{ulem}
\usepackage[dvipsnames]{xcolor}
\usepackage{booktabs}
\usepackage{makecell}
\begin{document}

\title{Emergent altermagnetism in (LaVO$_3$)$_n$/(LaNiO$_3$)$_m$ superlattices from interfacial charge transfer and octahedral tilts}

\author{Gustavo H. Cassemiro}
\affiliation{Departamento de Física, Universidade Federal de Minas Gerais, 
C. P. 702, 30123-970, Belo Horizonte, MG, Brazil}
\affiliation{Instituto de F\'{i}sica, Universidade de S\~ao Paulo, 05508-090 São Paulo, Brazil}
\author{Walber H. Brito}
\affiliation{Instituto de F\'{i}sica, Universidade de S\~ao Paulo, 05508-090 São Paulo, Brazil}
\email{walberbrito@usp.br}

\date{\today}

\begin{abstract}
Using first-principles calculations, we show that altermagnetism emerges in (LaVO$_3$)$_n$/(LaNiO$_3$)$_m$ superlattices, with $(n,m) = (1,1), (2,2)$ and $(2,4)$, from the interplay of interfacial charge transfer and octahedral tilts. Charge transfer drives the system towards the V$^{4+}$ and Ni$^{2+}$ configuration, turning the otherwise paramagnetic nickelate layers into a magnetically active sublattice. The resulting $C$-type antiferromagnetic order, combined with $a^-b^+c^-$ or $a^-b^-c^-$ octahedral tilts, links opposite-spin sublattices only through nonsymmorphic operations, yielding a $d$-wave spin splitting without spin-orbit coupling. Within DFT+U, the splitting reaches 118 meV in the insulating (LVO)$_1$/(LNO)$_1$ superlattice, about twice that calculated for bulk LaVO$_3$, and decreases as the nickelate layers become thicker.  Our findings indicate that this splitting is hosted by both the V and Ni sublattices, with nearly half of its weight on oxygen $2p$ states through $p$-$d$ covalency. Interfacial charge transfer thus emerges as a \textit{design} principle for engineering altermagnetism in oxide heterostructures.
\end{abstract}

\maketitle

\section{\label{sec:intro}INTRODUCTION}

Symmetry breaking and electronic correlations are among the most effective organizing principles in condensed matter physics, giving rise to a wealth of emergent electronic and magnetic phases in quantum  materials. For instance, one can mention unconventional superconductivity, charge and orbital order, and several forms of magnetism~\cite{imada1998metal}. Transition metal oxides are very promising materials in this field, as the strong interplay between charge, spin, orbital, and lattice degrees of freedom makes their ground states highly susceptible to small perturbations, allowing stabilization of distinct phases~\cite{Tokura_Science2000, Hwang2012}. A recent magnetic phase, namely altermagnetism, is a phase that combines the vanishing net magnetization, such as in a N\'eel type antiferromagnetic, with a momentum-dependent non-relativistic spin splitting arising when opposite spin sublattices are related by rotational rather than translational or inversion symmetries~\cite{vsmejkal2022beyond,vsmejkal2022emerging, mazin_altermag2022}. This spin splitting, tunable and present even without spin-orbit coupling (SOC), opens opportunities for the engineering of new spintronic devices~\cite{Jungwirth2026, Bai_AdvFunc_2026}, spin-charge conversion~\cite{Cui_2023PRB}, and topological phenomena~\cite{magnon_topologyPRB,Guerrero-Sanchez2026}.

Among the candidates, perovskite oxides are particularly attractive due to the interplay between their intrinsic octahedral tilts and orbital ordering, which can provide the symmetry-breaking mechanisms required to lift the spin degeneracy~\cite{naka2025altermagnetic, spaldin1_PRX2024,picozzi_PRB2025}. In fact, cooperative octahedral rotations of the GdFeO$_3$-type have been identified as a primary structural source of nonrelativistic spin splitting in several antiferromagnetic perovskites, and a list of bulk oxide candidates has been proposed~\cite{naka2025altermagnetic,Cano_2025JAP,Ganguli_PRB2025}, including vanadates and Ruddlesden-Popper nickelates. This indicates that the perovskite family is a fertile source of materials for altermagnetism, though this novel phase in these candidates appears as an intrinsic property of the bulk compounds.

Meanwhile, oxide superlattices (SLs) have emerged as a new route for engineering electronic and magnetic states. The accompanied electronic reconstruction redistributes the charge between the layers, and this interfacial charge transfer acts as a tuning knob to induce metal-insulator transitions, orbital reconstruction, and novel magnetic phases~\cite{Hwang2012,Chakhalian_RMP2014, zhong2017band,boris2011dimensionality, asmara2024emergence, beck2023interfacial}. Lanthanum nickelate (LaNiO$_3$) is a promising platform, where dimensional confinement and interfacial charge transfer reconstruct the electronic structure and can stabilize magnetic order in a paramagnetic correlated metal~\cite{catalano2018rare, boris2011dimensionality}. Charge transfer thus offers a new tunable direction that cannot be achieved in intrinsic bulk without introducing chemical disorder.

The search and realization of altermagnetic phases has so far been largely proposed based on theoretical works, with many candidates identified through symmetry screening and first principles calculations~\cite{vsmejkal2022emerging, naka2025altermagnetic}. On the other hand, oxide heterostructures are especially promising at this point, as they can be grown with atomic-layer precision using advanced methods, such as pulsed laser deposition (PLD) or molecular beam epitaxy (MBE)~\cite{Hwang2012, Chakhalian_RMP2014}. Yet, the deliberate \textit{design} of altermagnetism through interfacial charge transfer has remained largely unexplored.

In this work, using first-principles calculations, we show that altermagnetism can be realized at the interface of (LaVO$_3$)$_n$/(LaNiO$_3$)$_m$ superlattices. We find that an oxygen-continuity band mismatch drives charge transfer from LVO to LNO, driving the system towards a V$^{4+}$ (3d$^1$) and Ni$^{2+}$ (3d$^8$) configuration. The resulting $C$-type antiferromagnetic order, combined with cooperative octahedral tilts, connects the opposite spin sublattices through nonsymmorphic screw and glide operations, enforcing a $d$-wave spin splitting of up to 117.9 meV in the absence of SOC within a DFT+U insulating ground state. Orbital resolved band structures show that both the vanadium and nickel sublattices host the splitting, establishing the observed altermagnetism as a collective, charge transfer driven phase, distinct from a passive proximity effect. Our findings identify interfacial charge transfer as a \textit{design} principle for engineering altermagnetic phases in oxide superlattices.

\section{\label{sec:meth} COMPUTATIONAL METHODS}

Our density functional theory~\cite{DFT1, kohn1965self} calculations were performed within the revised Perdew-Burke-Ernzerhof generalized gradient approximation (PBESol)~\cite{pbesol_perdew}, using projector augmented wave (PAW) potentials~\cite{blochl1994projector}, as implemented in the Vienna Ab initio Simulation Package (VASP)~\cite{vasp1,vasp2}.
An energy cutoff of 600 eV was employed for the plane wave basis set, and the total energies were calculated with $\Gamma$-centered k-point meshes of $10 \times 8 \times 10$ in the Monkhorst-Pack scheme \cite{monkhorst1976special}. The total energy self consistent loops converged below $1\times10^{-7}$ eV. 
Our (LaVO$_3$)$_n$/(LaNiO$_3$)$_m$ supercells have dimensions $\sqrt{2} a_{pc} \times (n+m) a_{pc} \times \sqrt{2}a_{pc}$, with the LaVO$_3$ and LaNiO$_3$ layers alternating along the $b$ axis.  The (LVO)$_1$/(LNO)$_1$ SL was built from the orthorhombic $Pnma$ structure of LaVO$_3$ while the (LVO)$_2$/(LNO)$_2$ and (LVO)$_2$/(LNO)$_4$ from its monoclinic low-temperature structure. All the atomic positions and cell parameters were fully relaxed for each collinear magnetic configuration until the residual forces were smaller than 0.001 eV/\AA{}. Only symmetries that preserve the initial magnetic configuration were used during relaxation. 
Correlation effects were treated at a static mean field level, within the DFT+U approximation.
In particular, we used the rotationally invariant scheme proposed by Liechtenstein~\textit{et al.}~\cite{liechtenstein1995density}, where the correlation effects were included by on-site Coulomb repulsion $U$ and the Hund's coupling $J$. The adopted parameters were $U_V = 3.8$ eV and $U_{Ni} = 4.8$ eV, with $J = 0.8$ eV for both elements, corresponding to ($U-J_{H}$) of $3.0$ eV and $4.0$ eV for V and Ni, respectively, comparable to the $U_{eff}$ values used in previous works~\cite{Vanad_PRBUeff1,Ni_PRBUeff1}.
Different $U$ values were used to cross-check the stability of the calculated magnetic ground state for (LaVO$_3$)$_1$/(LaNiO$_3$)$_1$ as described in the Supplementary Material (Sec. I).

Our tight-binding (TB) model were obtained using maximally localized Wannier functions (MLWFs)~\cite{MLWFs_marzari} starting from the spin-polarized calculations. In particular, we wannierized each spin channel by considering the V-$3d$, Ni-$3d$ and the O-$2p$ orbitals of all atoms using Wannier90~\cite{wannier90}. As a result, we obtained a $pd$-model Hamiltonian which explicitly retains the O-$2p$ states, allowing us to resolve the ligand contribution to the spin splitting. 

\section{\label{sec:results}RESULTS AND DISCUSSIONS}

\subsection{Crystal symmetry analysis and altermagnetic order}

We first investigate the relaxed structures of (LaVO$_3$)$_n$/(LaNiO$_3$)$_{m}$ superlattices, with $(n,m) = (1,1), (2,2)$ and $(2,4)$; and their magnetic ground states.
To determine the magnetic ground state of each superlattice, we performed full structural relaxations considering various collinear magnetic configurations, namely $A$, $C$ and $G$-type antiferromagnetic (AFM) orders, as well as a ferromagnetic (FM) state. We have not imposed epitaxial strain conditions in our analysis.

\begin{figure}[!htb]
    \centering
    \includegraphics[scale=1]{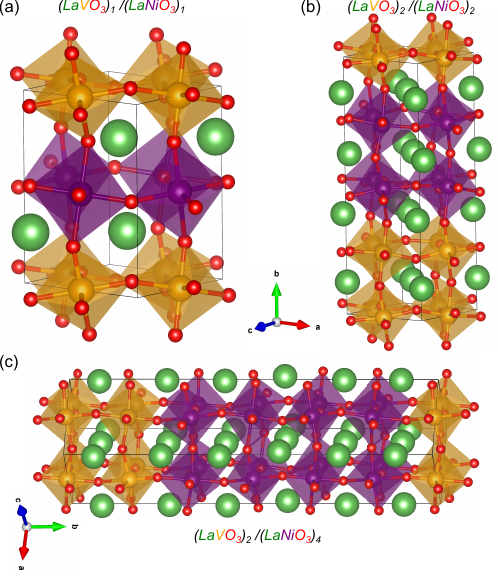}
    \caption{\textbf{Relaxed crystalline structures of the studied (LVO)$_n$/(LNO)$_m$ superlattices.} Schematic view of the (a) (LVO)$_1$/(LNO)$_1$ (space group  $P2_1/c$), (b) (LVO)$_2$/(LNO)$_2$ (space group $P_c$), and (c) (LVO)$_2$/(LNO)$_4$ (space group $P_c$) superlattices. The supercells are delimited by black solid lines, and the basis vectors \textbf{a} (red), \textbf{b} (green) and \textbf{c} (blue) are indicated by red, green, and blue arrows, respectively. The green, orange, purple, and red spheres represent La, V, Ni, and O atoms, respectively.}
    \label{fig:heterostructure_structural}
\end{figure}

The fully relaxed crystal structures of (LVO)$_1$/(LNO)$_1$,  (LVO)$_2$/(LNO)$_2$, and (LVO)$_2$/(LNO)$_4$  superlattices are shown in Figs. \ref{fig:heterostructure_structural} (a), (b), and (c), respectively.  
To clarify in more detail the structural distortions of the relaxed structures, we performed a symmetry analysis of each relaxed superlattice structure using the \textit{spglib} library~\cite{spglib}, with a tolerance of $10^{-3}$ \AA{}. In Glazer notation, we find that the oxygen octahedra in (LVO)$_1$/(LNO)$_1$, (LVO)$_2$/(LNO)$_2$, and (LVO)$_2$/(LNO)$_4$ are characterized by $a^-b^+c^-$,  $a^-b^+c^-$, and $a^-b^-c^-$ patterns, respectively.
We emphasize that this type of octahedral rotations is distinct from the one found in bulk LaNiO$_3$, which exhibits a rhombohedral structure with space group $R\overline{3}c$ and is classified as $a^-a^-a^-$ in Glazer notation~\cite{lnobulk}. In the case of LaVO$_3$, at high temperatures it crystallizes in an orthorhombic structure with space group $Pnma$ with octahedral rotations denoted as $a^-b^+a^-$~\cite{lvobulk}. At low temperatures it crystallizes in a monoclinic structure with space group $P2_1/a$ with octahedral rotations denoted as $a^-b^+c^-$~\cite{anas2023structural}. 

While the (LVO)$_1$/(LNO)$_1$ SL crystallizes in the centrosymmetric $P2_1/c$ space group,  the (LVO)$_2$/(LNO)$_2$ and (LVO)$_2$/(LNO)$_4$ SLs, built from the monoclinic structure of LaVO$_3$, relax to the non-centrosymmetric $Pc$ space group.
Throughout this work we use the standard space-group symbols, while the structures are described in a nonstandard setting with the unique monoclinic
axis along $\mathbf{c}$ and the layers stacked along $\mathbf{b}$. In this setting, the nonsymmorphic operations are a twofold screw axis along $\mathbf{c}$, $\{2_z\,|\,\tfrac{1}{2}\,0\,\tfrac{1}{2}\}$, and a glide plane perpendicular to $\mathbf{c}$ with intrinsic translation $\mathbf{a}/2$, $\{m_z\,|\,\tfrac{1}{2}\,0\,\tfrac{1}{2}\}$, which corresponds to the $c$-glide of the standard setting.
In the $Pc$ symmetry of the (LVO)$_2$/(LNO)$_2$ and (LVO)$_2$/(LNO)$_4$ SLs, the inversion center is lost, but the non-symmorphic glide plane operation survives. 

\begin{figure}[!htb]
    \centering
    \includegraphics[scale=1]{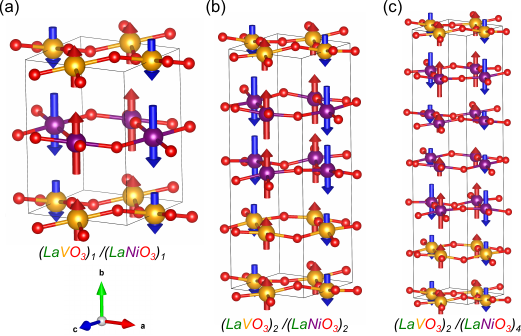}
    \caption{\textbf{$C$-AFM configurations of the (LVO)$_n$/(LNO)$_m$ superlattices.} Schematic view of the relaxed $C$-type antiferromagnetic structures of the (a) (LVO)$_1$/(LNO)$_1$, (b) (LVO)$_2$/(LNO)$_2$, and (c) (LVO)$_2$/(LNO)$_4$ superlattices. The supercells are delimited by black solid lines, and the basis vectors \textbf{a}, \textbf{b} and \textbf{c} are indicated by arrows. The orange, purple, and red spheres represent V, Ni, and O atoms, respectively. The La atoms were removed for better visualization. Local magnetic moments are represented by arrows on the transition metal sites.}
    \label{fig:mag_structure}
\end{figure}

Regarding the converged magnetic ordering, our total energy minimization establishes the $C$-AFM order as the ground state for (LVO)$_1$/(LNO)$_1$ and (LVO)$_2$/(LNO)$_2$, and (LVO)$_2$/(LNO)$_4$ SLs. These magnetic orderings are shown schematically in Figs.~\ref{fig:mag_structure} (a), (b), and (c), where the local moments are depicted as arrows.
To quantify the energetic stability of this emergent phase, we evaluated the relative energy difference per unit formula ($\Delta E_m = E_{m} - E_{C-AFM}$) between competing collinear magnetic configurations obtained in our calculations. Our results are shown in Table~\ref{tab:mag_stability}.

\begin{table}[!htb]
\caption{\textbf{Energetic stability of the collinear magnetic configurations.} Values are given as $\Delta E_m = E_m - E_{C\text{-AFM}}$ (meV/f.u.) for the (LVO)$_1$/(LNO)$_1$ ($P2_1/c$), (LVO)$_2$/(LNO)$_2$ $(P_c)$, and (LVO)$_2$/(LNO)$_4$ $(P_c)$
superlattices, where f.u.\ denotes one ABO$_3$ perovskite chemical unit. The $C$-AFM order consists of an in-plane checkerboard arrangement of antiparallel moments
with ferromagnetic stacking along $\mathbf{b}$ (as shown in Fig.~\ref{fig:mag_structure}).}
\begin{ruledtabular}
\begin{tabular}{lcccccc}
SL  & f.u./cell & $C$-AFM & $A$-AFM & $G$-AFM & FM  \\
\hline
(LVO)$_1$/(LNO)$_1$ & 4  & 0.0 & 58.2 & 6.9 & 50.3\\
(LVO)$_2$/(LNO)$_2$ & 8  & 0.0 & 67.7 & 98.5 & 73.6  \\
(LVO)$_2$/(LNO)$_4$ & 12 & 0.0 & 10.9 & 7.5 & 14.6 \\
\end{tabular}
\end{ruledtabular}
\label{tab:mag_stability}
\end{table}

For the (LVO)$_1$/(LNO)$_1$ SL, we find that the second lowest magnetic ordering is the G-AFM configuration, lying 6.9 meV/f.u. above the ground state.
In the (LVO)$_2$/(LNO)$_2$ SL, the second lowest energy state is the $A$-AFM order, followed by the FM configuration. Finally, for the (LVO)$_2$/(LNO)$_4$ SL we observe that the $G$-type AFM is once again the second most stable ordering, as observed in the (LVO)$_1$/(LNO)$_1$ SL.
In addition, we systematically varied the Hubbard parameters (U$_{Ni}$ and U$_{V}$) for the (LVO)$_1$/(LNO)$_1$ superlattice. 
The $C$-AFM phase remains the robust magnetic ground state considered in this work, for sizable U values. This suggests that the stabilization of the $C$-AFM order is not a spurious artifact of the chosen correlation parameters, but the intrinsic driven response of the system to the new electronic structure. 

More importantly, the obtained relaxed structure and magnetic configurations suggest that the transition metal atoms cannot be mapped onto one another by simple lattice translations. Despite this lowering of the symmetry, both structural phases host the nonsymmorphic structural symmetries; this specific interplay with the $C$-AFM collinear order provides the basic ingredients for the emergence of altermagnetism~\cite{vsmejkal2022beyond,vsmejkal2022emerging}. 
To disclose the putative altermagnetic character of the spin splitting based on symmetry analysis, we investigate the spin-group operations that relate the two magnetic sublattices (within the same chemical Ni and V species) for our investigated structures. In particular, we obtained the space group of each fully relaxed structure and magnetic ground state. For every symmetry operation, we evaluate the induced permutation of the transition metal sites and compare the sign of the calculated local moment before and after operation. For each system, we obtain the space group and spin swapping and preserving  operations listed in Table~\ref{tab:symmetries_mag_struc}.

\begin{table}[htbp]
    \centering
    \renewcommand{\arraystretch}{1.2}
    \setlength{\tabcolsep}{4pt}
    \caption{\textbf{Symmetry operations of the magnetic superlattices.} Space groups (standard symbols) of the relaxed superlattices (SL) and the spin-swapping (SS) and spin-preserving (SP) symmetry operations of their $C$-AFM ground states. Operations are given in Seitz notation $\{R\,|\,\mathbf{t}\}$ for the setting used in this work (unique monoclinic axis along $\mathbf{c}$, layers stacked along $\mathbf{b}$). The twofold screw axis along $\mathbf{c}$ is denoted by $\{2_z\,|\,\tfrac12\,0\,\tfrac12\}$ and the glide plane perpendicular to the $\mathbf{c}$ axis is denoted by $\{m_z\,|\,\tfrac12\,0\,\tfrac12\}$, which corresponds to the $c$-glide of the standard setting.}
    \resizebox{\columnwidth}{!}{%
        \begin{tabular}{@{}ccll@{}}
            \toprule
            SL & Space group & SS & SP \\
            \midrule
            (LVO)$_1$/(LNO)$_1$ & $P2_1/c$
                & $\{2_z\,|\,\tfrac12\,0\,\tfrac12\}$, $\{m_z\,|\,\tfrac12\,0\,\tfrac12\}$
                & $\{E\,|\,0\}$, $\{\bar{1}\,|\,0\}$ \\
            (LVO)$_2$/(LNO)$_2$ & $Pc$
                & $\{m_z\,|\,\tfrac12\,0\,\tfrac12\}$ & $\{E\,|\,0\}$ \\
            (LVO)$_2$/(LNO)$_4$ & $Pc$
                & $\{m_z\,|\,\tfrac12\,0\,\tfrac12\}$ & $\{E\,|\,0\}$ \\
            \bottomrule
        \end{tabular}
    }
    \label{tab:symmetries_mag_struc}
\end{table}

For the (LVO)$_1$/(LNO)$_1$ superlattice, the relaxed structure has two V sites which are crystallographically equivalent (Wyckoff $c$), as are the two Ni sites (Wyckoff $d$), and their obtained local moments are the same in magnitude, $|m_V| = 1.01 \mu_B$ for each V atom and $|m_{Ni}| = 1.63 \mu_B$ for each Ni atom. Of the four operations, the identity ($E$) and the inversion center are spin preserving (SP), while the $2_1$ screw axis and the glide plane are spin-swapping (SS). We emphasize that neither spin-swapping operation is a pure translation or an inversion; therefore, the opposite spin sublattices are linked solely by the nonsymmorphic screw and glide operations. Therefore, our symmetry analysis suggests that our SLs can host an altermagnetic phase with symmetry-allowed spin splitting characterized by a $d$-wave form factor~\cite{vsmejkal2022beyond, roigPRBMinimalModel}. We note that $G$-AFM configuration, which lies 6.9 meV/f.u. above the $C$-AFM ground state, shares the same spin group, with $2_1$ screw axis and the glide plane also exchange its opposite-spin sublattices. The putative altermagnetic phase for (LVO)$_1$/(LNO)$_1$ is therefore robust against the competition between the two lowest-energy magnetic configurations.

Moreover, increasing the layer thickness reduces the symmetry. Both the (LVO)$_2$/(LNO)$_2$ and (LVO)$_2$/(LNO)$_4$ superlattices relax to non-centrosymmetric space group $P_c$. As a result, in these structures the glide plane is the only nonsymmorphic operation. According to our analysis, the glide plane continues to connect the opposite spin sublattices without any translation or inversion, indicating that the altermagnetic classification ($d$-wave type) is preserved.  

Finally, we emphasize that recent studies on altermagnetic perovskites have explicitly demonstrated that specific cooperative octahedral tiltings are the primary structural mechanism for lifting the non relativistic spin degeneracy in these systems~\cite{naka2025altermagnetic}. This symmetry configuration strictly dictates that the lifts of spin degeneracy are momentum-dependent even in the absence of spin-orbit coupling (SOC).

\subsection{Charge transfer and magnetic moments}
\label{subsec:chargetransfer}

To clarify the physical mechanism responsible for the stabilization of a $C$-AFM ground state in our SLs, we next study the charge transfer between both perovskites. In particular, we applied the oxygen continuity boundary condition, which allows us to predict the induced charge transfer between the superlattice constituents starting from the electronic structure of the bulk compounds~\cite{zhong2017band}. 

We evaluated the projected density of states (PDOS) of isolated bulk materials in their converged structural and magnetic ground states. 
Our bulk calculations stabilize the $C$-type antiferromagnetic order in LaVO$_3$ with a monoclinic $P2_1/a$ structure, which is the low-temperature phase of LVO. This is consistent with experimental findings of the low-temperature phase, in which the symmetry lowering from orthorhombic $Pbnm$ to monoclinic $P2_1/b\,1\,1$ is associated with $G$-type orbital ordering~\cite{LVO_exp1}, while neutron diffraction experiments confirm $C$-type AFM order~\cite{LVO_exp2}.
On the other hand, for bulk LaNiO$_3$, the application of a finite Hubbard $U_{eff}$ to capture strong correlations artificially penalizes hybridization, suppressing its intrinsic paramagnetic metallic phase locking the system into a ferromagnetic  (FM) ground state, although previous studies demonstrate that smaller values of $U_{eff}$ can capture the AFM ground state of LaNiO$_3$ \cite{mercy2017structurally,varignon2017complete,catalano2018rare}. Despite this known limitation of the DFT+U framework for bulk LNO, we used these converged states ($C$-AFM for LVO and FM for LNO) as a baseline to estimate the interfacial charge transfer. 

\begin{figure}[!htb]
    \centering
    \includegraphics[scale=1.2]{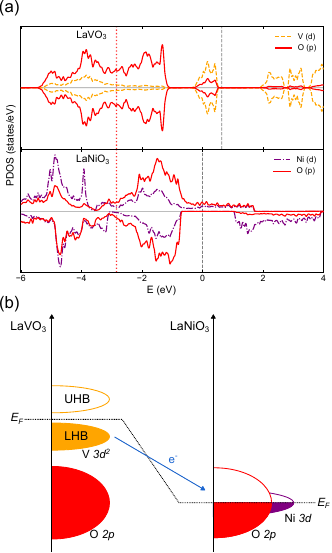}
    \caption{\textbf{Electronic structure of the constituent bulk materials.} (a) Spin-polarized projected density of states (PDOS) for isolated bulk LaVO$_3$ (top) and LaNiO$_3$ (bottom), aligned by their oxygen $2p$ band centers. The black vertical dashed line indicates the Fermi energy ($E_F$). (b) Schematic illustration of the corresponding bulk band alignments. According to the model of Ref.~\cite{zhong2017band}, the pronounced energy offset between the oxygen $2p$ band centers drives a spontaneous interfacial electron transfer from the V-$3d$ states to the Ni-$3d$-O $2p$ states upon heterostructure formation to equalize the electronic chemical potential.}   
    \label{fig:pdos_bulks_scheme}
\end{figure}

In Fig.~\ref{fig:pdos_bulks_scheme} (a) we present the calculated spin-polarized projected density of states (PDOS) of the bulk materials where there is a sizable energy mismatch between the O-$2p$ electronic states of LaVO$_3$ and LaNiO$_3$. Our calculations show that the center of gravity of the oxygen $2p$ bands in LaVO$_3$ is located at $-3.48$ eV relative to the Fermi energy, lying deeper than the corresponding oxygen states in LaNiO$_3$, that is $-2.84$ eV. This offset of approximately $0.64$ eV reflects the distinct local charge-transfer energies $\epsilon_{dp}$ of the two bulks systems. Following the oxygen continuity condition proposed in Ref.~\cite{zhong2017band}, at the interface, the oxygen $2p$ states need to be align across the interface upon the formation of the superlattice. As reported for complex oxide interfaces, the component with the lower $\epsilon_p$ acts as the intrinsic electron donor \cite{zhong2017band}. Therefore, in order to maintain the constant E$_F$ throughout the superlattice in equilibrium, the system must undergo an electronic reconstruction. The mismatch ($\Delta\epsilon_p$) of $0.64$ eV acts as the main driving force for spontaneous and directional charge transfer, indicating that electrons are transferred from the V-$3d$ orbitals to hybridized Ni-$3d$ - O-$2p$ states, as schematically shown in Fig.~\ref{fig:pdos_bulks_scheme}(b). 

\begin{figure}[!htb]
    \centering
    \includegraphics[scale=0.94]{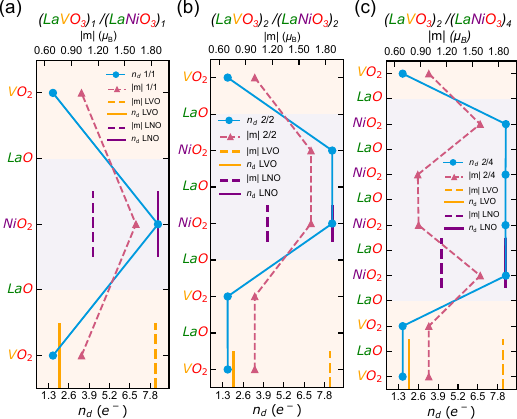}
    \caption{\textbf{Layer resolved charge transfer and induced local magnetic moments in the superlattices.} Layer-resolved $d$ electron occupancy $n_d$ (only $t_{2g}$ for V, bottom axis, blue circles) and magnetic moment magnitude $|m|$ (top axis, pink triangles) for the (a) (LaVO$_3$)$_1$/(LaNiO$_3$)$_1$, (b) (LaVO$_3$)$_2$/(LaNiO$_3$)$_2$, and (c) (LaVO$_3$)$_2$/(LaNiO$_3$)$_4$ superlattices. Vertical lines denote bulk reference values for LaVO$_3$ (orange) and LaNiO$_3$ (purple).}
    \label{fig:mag_occup_SL}
\end{figure}

In fact, by integrating the projected density of states up to the Fermi energy, we find that the V-t$_{2g}$ occupancies $(n_d)$ in our SL's are reduced compared to its bulk reference. To quantify the spatial extent of this charge transfer, we evaluate the layer resolved $d$-orbital  occupancy $n_d$ and local magnetic moments $|m_V|$ for both the V and Ni sites across the SL's. Our results are shown in Fig.~\ref{fig:mag_occup_SL}. In these plots, the occupancies are indicated by continuous lines and the local magnetic moments are indicated by dashed lines. The bulk references values are indicated by vertical lines, where the orange lines represent the pristine bulk LaVO$_3$, while purple lines represent the LaNiO$_3$ bulk. As can be noticed in Fig.~\ref{fig:mag_occup_SL}(a), for the (LVO)$_1$/(LNO)$_1$ SL we observe that V occupancy ($n_d$) drops by $-0.41$ e$^-$/atom below its bulk reference, forcing the V cation out of its nominal $3d^2$ Mott-insulating state towards a $3d^1$-like configuration. This depletion in charge is related to the calculated local moments, where the magnetic moment of vanadium in the superlattice is $1.01$ $\mu_B$/V atom, which favors the emergent $C$-AFM interfacial order in the superlattice.

Here it is important to understand in more details the effects of the charge transfer and  structural distortions for the stabilization of the $C$-AFM phase.
For the (LVO)$_1$/(LNO)$_1$ superlattice, we performed control calculations in which all V atoms in the structure  were replaced by Ni, while maintaining the lattice parameters and atomic positions (see Supplementary Material Sec. II). This hypothetical LaNiO$_3$ structure ($P2_1/c$) retains the structural distortions presented in the relaxed (LVO)$_1$/(LNO)$_1$ superlattice, allowing us to investigate whether these distortions alone favor the $C$-AFM order. Among the converged magnetic configurations, the FM state has the lowest energy, whereas the $A$-AFM, $C$-AFM, and $G$-AFM configurations lie 42.6, 142.1, and 339.1 meV/f.u. higher in energy, respectively. Thus, imposing the superlattice geometry on LaNiO$_3$ is insufficient to reproduce the $C$-AFM ground state found in the (LVO)$_1$/(LNO)$_1$. This comparison supports the importance of the chemical and electronic reconstruction introduced by the LVO layers, including interfacial charge transfer, which helps to stabilize the $C$-AFM order. However, because the substitution also modifies the local electronic structure and magnetic exchange interactions, this control calculation does not isolate the contribution of charge transfer alone.

Moreover, in our (LVO)$_1$/(LNO)$_1$ SL we observe that Ni $n_d$ occupancy essentially does not change relative to its bulk reference (vertical continuous purple line), indicating that the transferred charge is accommodated in the nickelate layers in a fundamentally different way. Bulk LaNiO$_3$ is well established as a negative charge-transfer metal, in which the nominal Ni$^{2+}$ valence is realized as a strongly hybridized $3d^8\underline{L}$ ground state, where $\underline{L}$ denotes a hole in the O-$2p$ ligand orbitals~\cite{catalano2018rare}. The electrons flowing from the vanadate layers therefore primarily fill these ligand-hole states rather than directly increasing the Ni-$3d$ occupancy, which explains why the projected Ni occupancy remains essentially unchanged despite the net influx of charge. The result is a ligand-hole free Ni$^{2+}$ (3d$^8$, $S=1$) configuration. We stress that quenching the ligand holes does not remove the Ni-O covalency. The Ni $e_g$ - O-$2p$ hybridization remains strong in the 3$d^8$ state, as reflected in the obtained reduced Ni local moment of $1.63 \mu_B$, compared with the ionic value of $2 \mu_B$ for $S=1$, whose remaining spin density is transferred to the surrounding oxygens.

To establish the spatial extent of the charge transfer and induced local moments to the thicker superlattices, we expand our analysis to the (LVO)$_2$/(LNO)$_2$ and (LVO)$_2$/(LNO)$_4$ systems. As shown in Figs.~\ref{fig:mag_occup_SL}(b) and (c), the layer resolved charge transfer and magnetic moments exhibit a spatial reduction away from the interface. While the interfacial Ni sites exhibit enhanced magnetic moments, the inner Ni sites show a reduction of magnetization (smaller than the LaNiO$_3$ bulk). This magnetic gradient shows that the transferred electrons from vanadate layers are screened within approximately one unit cell of the interface. This demonstrates that the uniform $3d^8$ phase observed in the (LVO)$_1$/(LNO)$_1$ and (LVO)$_2$/(LNO)$_2$ systems is strictly enabled in ultrathin limit. This spatial accommodation and local electronegativity equalization is further corroborated by the continuous spatial profile of the effective charge transfer energy ($\epsilon_{dp}$) across the superlattice (see Supplementary Material Sec. III).

The structural signature of this interfacial charge transfer is confirmed by the evolution of the local octahedral volumes (see Supplementary Material Sec. IV). 
On one hand, for NiO$_6$ octahedra within the same layer, we do not observe any breathing distortions, consistent with the Ni$^{2+}$ (3d$^8$) state, for which bond disproportionation is not expected. On the other hand, we observe a pronounced inter-layer volume shift. The VO$_6$ octahedra undergo a volumetric contraction of around 1.2 $\AA{}^3$ - 1.7 $\AA{}^3$, relative to the bulk values, consistent with a V$^{3+}$ to V$^{4+}$ valence shift. Simultaneously, the response of the NiO$_6$ octahedra is highly dependent on the dimensional confinement. In the fully confined (LaVO$_3$)$_1$/(LaNiO$_3$)$_1$ and (LaVO$_3$)$_2$/(LaNiO$_3$)$_2$, the entire nickelate layers are interfacial ones, so they uniformly absorbs the transferred electrons, resulting in a volume expansion of around 1.6 $\AA{}^3$ - 1.7 $\AA{}^3$, relative to the bulk values, quenching the ligand holes and driving the nickelate layers towards a homogeneous $3d^8$ (Ni$^{2+}$). In the thicker (LaVO$_3$)$_2$/(LaNiO$_3$)$_4$ SL, this volume expansion is suppressed in the inner bulk-like layers (around 0.3 $\AA{}^3$) compared to the interfacial planes (around 1.1 $\AA{}^3$). This spatial gradient indicates that the charge transfer is screened within a few unit cells. This behavior is similar to the interfacial electron doping observed in titanate-nickelate superlattices \cite{asmara2024emergence}.

As pointed out in recent literature~ \cite{naka2025altermagnetic}, while bulk LaVO$_3$ can host altermagnetism on a nominal background $3d^2$, the symmetry-breaking conditions in our heterostructures can contribute to the formation of the emergent altermagnetic phase from this novel interface-driven $3d^1$ like state.

\subsection{Electronic Structure and non-relativistic spin splittings}

\begin{figure}[!htb]
    \centering
    \includegraphics[scale=0.94]{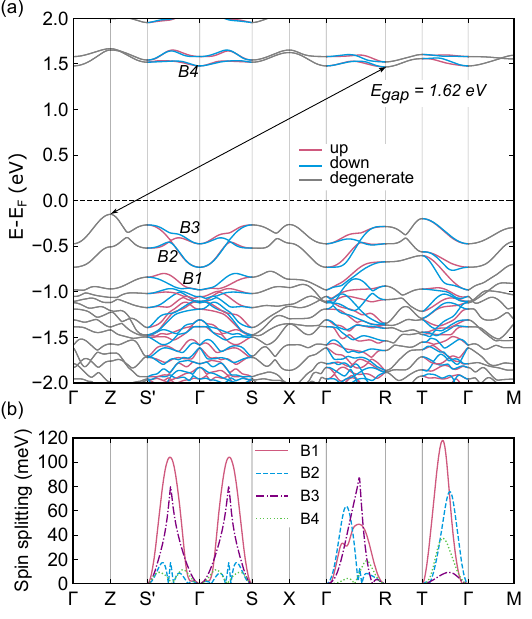}
    \caption{\textbf{Spin-resolved electronic band structure of the (LVO)$_1$/(LNO)$_1$ superlattice.} (a) Calculated DFT+U band structure for the $C$- AFM ground state. Spin-degenerate bands are colored in gray, while the spin splitting character are colored in red and blue for spin-up and spin-down, respectively, revealing the altermagnetic spin-splitting character. The top three valence bands and first conduction band are denoted by $B1$, $B2$, $B3$ and $B4$, respectively. (b) Momentum-resolved magnitude of the non-relativistic spin splitting $\Delta E(\mathbf{k})$ for each denoted band.}
    \label{fig:bands_LVO_LNO_1_1}
\end{figure}

\begin{figure}[!htb]
    \centering
    \includegraphics[scale=0.94]{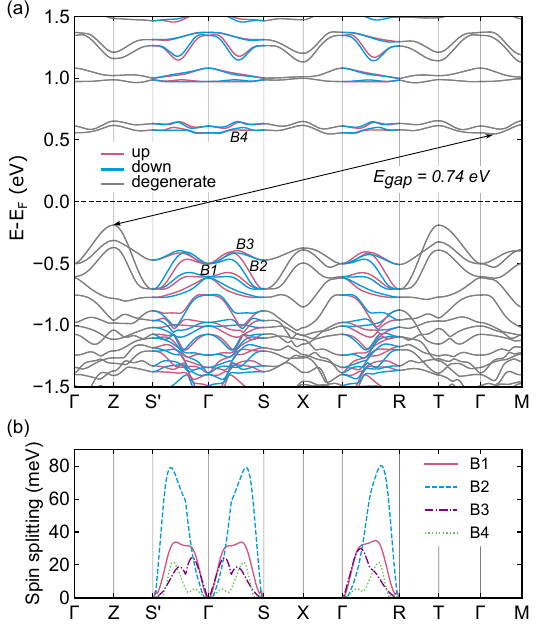}
    \caption{\textbf{Spin-resolved electronic band structure of the (LVO)$_2$/(LNO)$_2$ superlattice.}(a) Calculated DFT+U band structure for the $C$-type AFM state. Spin-degenerate bands are colored in gray, while the spin splitting character are colored in red and blue for spin-up and spin-down, respectively, revealing the altermagnetic spin-splitting character. The top three valence bands and first conduction band are denoted by $B1$, $B2$, $B3$ and $B4$, respectively. (b) Momentum-resolved magnitude of the non-relativistic spin splitting $\Delta E(\mathbf{k})$ for each denoted band.}
    \label{fig:bands_LVO_LNO_2_2}
\end{figure}

\begin{figure}[!htb]
    \centering
    \includegraphics[scale=0.94]{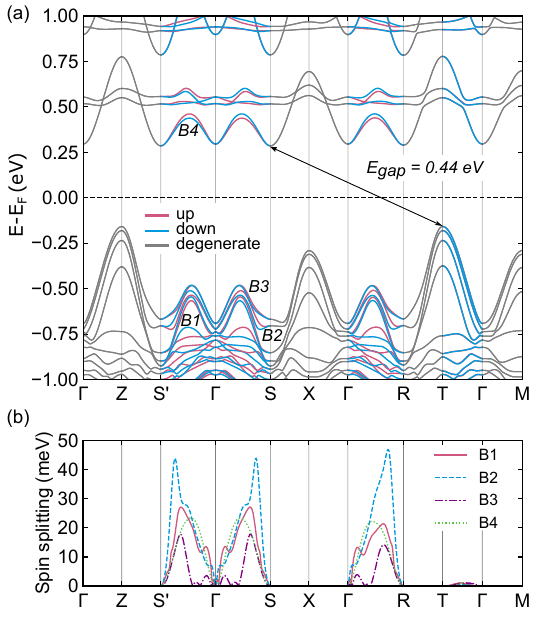}
    \caption{\textbf{Spin-resolved electronic band structure of the (LVO)$_2$/(LNO)$_4$ superlattice.} (a) Calculated DFT+U band structure sout{of (LVO)$_2$/(LNO)$_4$} for the $C$-type AFM state. Spin-degenerate bands are colored in gray, while the spin splitting character are colored in red and blue for spin-up and spin-down, respectively, revealing the altermagnetic spin-splitting character. The top three valence bands and first conduction band are denoted by $B1$, $B2$, $B3$ and $B4$, respectively. (b) Momentum-resolved magnitude of the non-relativistic spin splitting $\Delta E(\mathbf{k})$ for each denoted band.}
    \label{fig:bands_LVO_LNO_2_4}
\end{figure}

After describing the symmetry properties, magnetic configurations, and charge transfer, we next study the spin-resolved electronic structure of the SLs.
As can be seen in Figs.~\ref{fig:bands_LVO_LNO_1_1} (a), ~\ref{fig:bands_LVO_LNO_2_2} (a) and ~\ref{fig:bands_LVO_LNO_2_4} (a), the obtained band structures calculated without the inclusion of the spin-orbit coupling (SOC), demonstrate the existence of momentum-dependent spin splittings. In particular, the spin degeneracy is preserved for (LVO)$_1$/(LNO)$_1$ along $\Gamma-Z-S'$, $S-X-\Gamma$,  $R-T$ and $\Gamma-M$ directions (See Appendix~\ref{appendixA}). On the other hand, the band structures along the $S'-\Gamma-S$, $\Gamma - R$ and $T - \Gamma$ directions show the lift of the spin degeneracy, characteristic of an altermagnetic phase. 
Similar features are observed for the (LVO)$_2$/(LNO)$_2$ and (LVO)$_2$/(LNO)$_4$ SLs.
To quantify these non-relativistic spin splittings, we calculate the momentum resolved splitting magnitude of selected bands near the band edges (denoted by $B1$, $B2$, $B3$, and $B4$). Our results are shown in Figs.~\ref{fig:bands_LVO_LNO_1_1}(b),~\ref{fig:bands_LVO_LNO_2_2}(b), and ~\ref{fig:bands_LVO_LNO_2_4}(b).

For the (LVO)$_1$/(LNO)$_1$ SL, the band $B1$ reaches a maximum splitting of approximately $117.9$ meV along the $T - \Gamma$ high-symmetry direction. To contextualize this magnitude, we calculated the spin splitting of the pristine bulk LaVO$_3$ in its low temperature monoclinic phase (see Supplementary Material Sec. V). While bulk LaVO$_3$ hosts the altermagnetism, its maximum spin splitting reaches  $\approx 57.1$ meV. Therefore, the confinement of LVO within the (LVO)$_1$/(LNO)$_1$ SL enhances the altermagnetic spin splitting for bands around the edges. This enhancement is a consequence of the confinement at the interface and the maximization of the out-of-plane V-O-Ni apical angles that promotes a strongly anisotropic inter orbital hoppings. As the layer thickness increases in the (LVO)$_2$/(LNO)$_2$ superlattice, this maximum splitting decreases to roughly $80.3$ meV, for band $B2$. In the (LVO)$_2$/(LNO)$_4$ superlattice, as the inner nickelate layers try to recover their non-magnetic bulk-like state, the maximum spin splitting further goes to approximately $47$ meV (band $B2$). This reduction suggests that the altermagnetic phase is linked to the interfacial charge reconstruction. The calculated non-relativistic splitting on the order of $\approx 100$ meV in the fully confined limit superlattice is comparable to the giant Rashba splittings of polar semiconductors such as the BiTeI~\cite{ishizaka2011giant}, and twice as large as the splitting of bulk LaVO$_3$ according to our calculations.
This finding further emphasizes the influence of the anisotropic crystal field generated by the $a^{-}b^{+}c^{-}$ octahedral tilt pattern.

Although the pristine bulk of LaNiO$_3$ is a well-known correlated paramagnetic metal, our superlattice results demonstrate that  confinement and interfacial structural distortions reconstruct its electronic and magnetic state. Drawing parallels to the confinement-driven metal-insulator transitions (MIT) extensively reported for correlated oxide superlattices and heterostructures, the spatial confinement of the (LVO)$_1$/(LNO)$_1$ superlattice drives the system into an insulating regime within DFT+U, with an indirect band gap of $1.62$ eV. However, as the thickness of the sublayer increases in the (LVO)$_2$/(LNO)$_2$ system, the DFT+U gap is significantly reduced to $0.74$ eV. This trend is confirmed in the (LVO)$_2$/(LNO)$_4$ superlattice, where the DFT+U gap reduces further to $0.44$ eV. This thickness dependence reflects the drive of the nickelate sublayers to recover their intrinsic metallic character as the dimensional confinement is gradually relaxed.

To investigate the role of the octahedral tilts, we performed additional control calculations for an idealized untilted (LVO)$_1$/(LNO)$_1$ SL with  $P4/mmm$ symmetry (see Supplementary Material Sec. II.). With $J=0.8$ eV, the FM configuration is the ground state, and all initial AFM configurations relax into ferrimagnetic states with incomplete compensated moments. With $J=0$, the FM state remains the ground state, and the initial $C$-AFM configuration converges into the $G$-AFM state.

The energetics of the relaxed tilted structure clarifies how the $C$-AFM order is selected. In the (LVO)$_1$/(LNO)$_1$ SL, the $C$-AFM and $G$-AFM configurations share the same antiparallel in-plane arrangement of local moments and differ only in the alignment of the V and Ni moments across the interface, which is parallel in the $C$-AFM state, as shown in Fig.~\ref{fig:mag_structure}. In addition, the $C$-AFM and FM configurations share the same interfacial alignment and differ only within the planes. The large energy cost of the FM (50.3 meV/f.u.) and $A$-AFM (58.2 meV/f.u.) configurations therefore indicates a strong antiferromagnetic coupling within the NiO$_2$ and VO$_2$ planes. In contrast, the small difference between the $G$-AFM and $C$-AFM configurations (6.9 meV/f.u.) reflects a weak ferromagnetic coupling across the interface. These findings are consistent with the Goodenough-Kanamori-Anderson rules~\cite{GKA}. In particular, within the planes, the $e_g^{2}$ - $e_{g}^2$ (Ni-O-Ni) and $t_{2g}^1 - t_{2g}^1$ (V-O-V) superexchange is antiferromagnetic. Across the interface, the hopping from Ni $e_g$ states into the empy V $e_g$ states favors a ferromagnetic coupling, which selects the $C$-AFM over the $G$-AFM configuration.

In the untilted structure (control calculation), this local moment picture is strongly affected. The AFM configurations turn into ferrimagnetic states due to the instability of the $C$-AFM configuration. This indicates that, without the octahedral tilts, the larger bandwidth favors the magnetism to be itinerant like, favoring the FM state within DFT+U. The octahedral tilts thus play a dual role. From the energetic point of view, they narrow the bands and stabilize the local moments whose superexchange selects the $C$-AFM order. In addition, they remove the lattice translation symmetries that connect the two spin sublattices in the untilted structure. As a result, in the $G$-AFM state obtained for the idealized structure, there is no spin splitting, as shown in Fig.~S2 of Supplementary Material.

\begin{figure*}[!htb]
    \centering
    \includegraphics[scale=0.9]{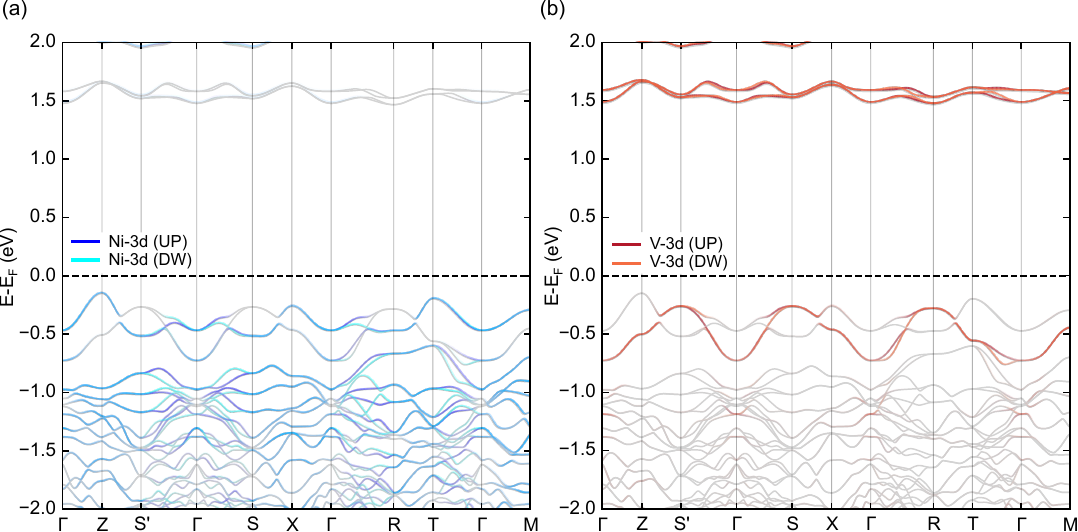}
    \caption{\textbf{Orbital-projected band structure for the (LVO)$_1$/(LNO)$_1$ superlattice.} The electronic band dispersion is projected onto the (a) Ni-$3d$ and (b) V-$3d$ states. Spin-up (UP) and spin-down (DW) channels are distinguished by color.}
    \label{fig:pband_V_Ni_1_1}
\end{figure*}

To further elucidate the orbital origin and layer dependence of low-energy bands and their spin-splittings, we projected the band structures onto the local TM (Ni,V) 3$d$ orbitals. In Fig.~\ref{fig:pband_V_Ni_1_1} we show the orbital resolved bands for the (LVO)$_1$/(LNO)$_1$ SL. The orbital-resolved dispersion demonstrates that the momentum-dependent spin splitting is not confined to a single sublayer. For instance, along the $\Gamma-S$, $\Gamma-R$ and $T-\Gamma$ paths, both the V-$3d$ and Ni-$3d$ projected states exhibit an active spin-degeneracy lifting. This direct visualization is essential to distinguish the fundamental nature of our superlattice from recent proposals in the field. Recently, it has been demonstrated that momentum-alternating spin-splitting can be transferred across an interface to an adjacent nonmagnetic layer via the altermagnetic proximity effect (AMPE)~\cite{zhu2026altermagnetic}. In such proximity-driven platforms, such as PbO/V$_2$Se$_{2}$O, the nonmagnetic material acquires spin splitting passively through interfacial wavefunction hybridization without altering its intrinsic chemical state. In contrast, our projected band structure shows that the splitting in the (LVO)/(LNO) superlattices is not a passive proximity effect. Driven by the interfacial charge transfer that quenches oxygen ligand holes, Ni cations are fundamentally reconfigured to be in an active,  $3d^8$ state. Consequently, both the vanadate and nickelate layers actively host and participate in the newly established $C$-AFM lattice. 

\begin{figure*}[!htb]
    \centering
    \includegraphics[scale=0.9]{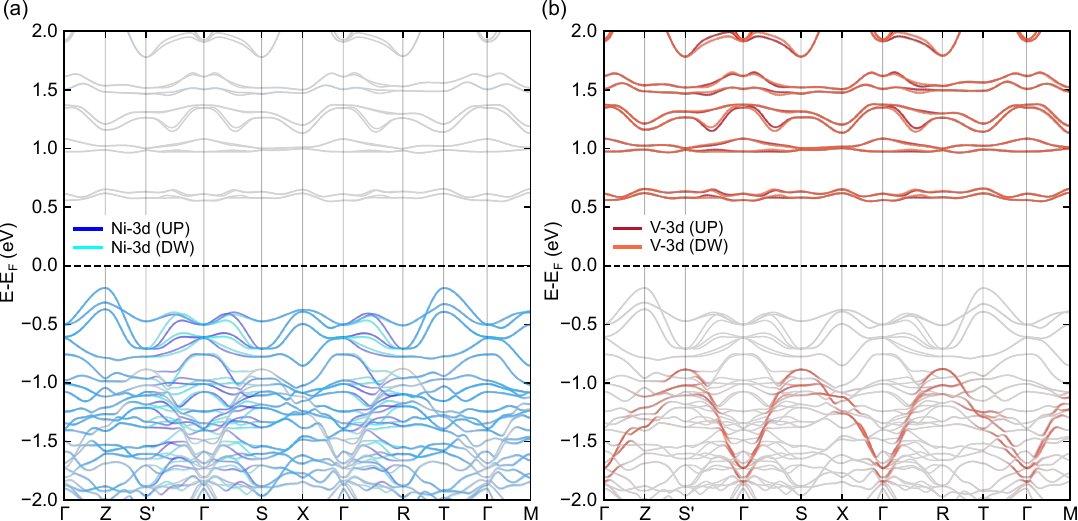}
    \caption{\textbf{Orbital-projected band structure for the (LVO)$_2$/(LNO)$_2$ superlattice.} The electronic band dispersion is projected onto the local (a) Ni-$3d$ and (b) V-$3d$ states. Spin-up (UP) and spin-down (DW) channels are distinguished by color.}
    \label{fig:pband_V_Ni_2_2}
\end{figure*}

\begin{figure*}[!htb]
    \centering
    \includegraphics[scale=1]{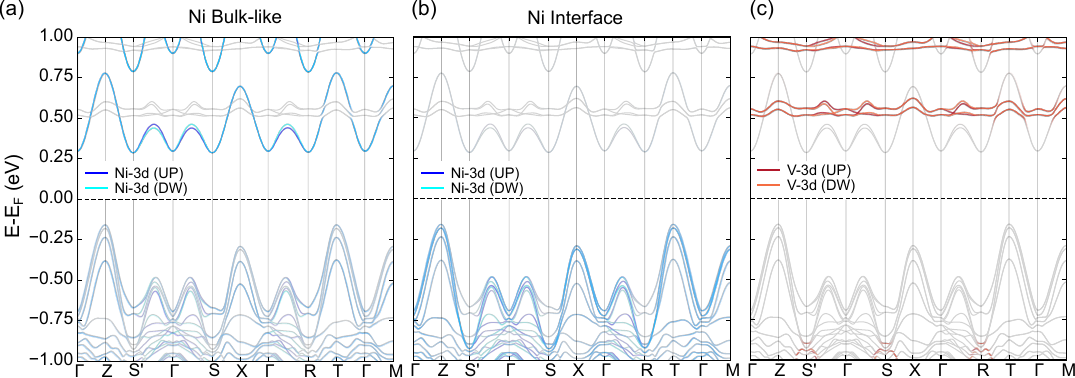}
    \caption{\textbf{Orbital-projected  band structure for the (LVO)$_2$/(LNO)$_4$ superlattice.} The electronic band structure is projected onto the (a) intra-layer (bulk-like) Ni-$3d$, (b) interfacial Ni-$3d$, and (c) V-$3d$ states. Spin-up (UP) and spin-down (DW) channels are distinguished by color.}
    \label{fig:pband_V_Ni_2_4}
\end{figure*}

For the (LVO)$_2$/(LNO)$_2$ SL, we observe that the top of the valence band is mostly composed of Ni-3$d$ states, which exhibit the spin-splittings (see Fig.~\ref{fig:pband_V_Ni_2_2}). In addition, as can be seen in Fig.~\ref{fig:pband_V_Ni_2_4} for the (LVO)$_2$/(LNO)$_4$ SL, we observe that the interfacial Ni-3$d$ states are responsible for the top of the valence band and are responsible for the spin-splittings observed in our band structure. The V-3$d$ states, in contrast, are responsible for the bottom of the conduction band, where it is possible to observe the spin-splittings. 

\subsection{\label{sec:wannier} Symmetry classification and microscopic origin of spin-splittings}

To address the microscopic origin of the observed momentum-dependent spin splittings and symmetry classification of the observed altermagnetic order, we constructed maximally localized Wannier functions (MLWFs)~\cite{MLWFs_marzari} for the collinear magnetic ground state ($C$-AFM) of the (LaVO$_3$)$_1$/(LaNiO$_3$)$_1$ superlattice (See Sec.~\ref{sec:meth} for additional details). The interpolated bands reproduce the DFT band structure closely within the frozen window (up to $\approx 3.5$ eV above $E_F$), as detailed in Supplementary Material Sec. VI.

\begin{figure*}[!htb]
    \centering
    \includegraphics[scale=0.8]{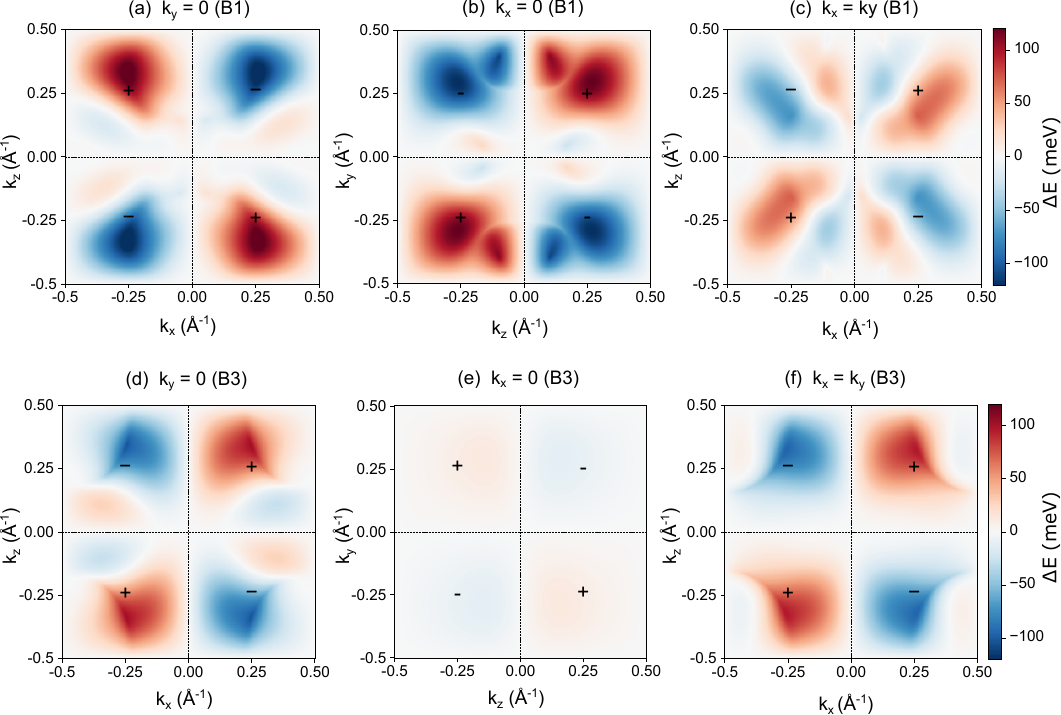}
    \caption{\textbf{Momentum-dependent spin splittings in the (LaVO$_3$)$_1$/(LaNiO$_3$)$_1$ superlattice.} Sign-resolved non-relativistic spin splitting $\Delta E(\mathbf{k})$ of the occupied band ($B1$) across the (a) $k_y=0$, (b) $k_x=0$, and diagonal (c) $k_x=k_y$ momentum planes. (d)-(f) Corresponding $\Delta E(\mathbf{k})$ contour plots for the $B3$ band. The distinct momentum-dependent alternation of positive (red, $+$) and negative (blue, $-$) lobes demonstrates the symmetry-enforced mixed $d$-wave ($d_{xz} + d_{yz}$) altermagnetic form factor imposed by the $P2_1/c$ symmetry.}
    \label{fig:altermag_charac}
\end{figure*}

Next, we diagonalize the TB Hamiltonians $H_\uparrow(\mathbf{k})$ and
$H_\downarrow(\mathbf{k})$ across the Brillouin zone, and find that
\begin{equation}
  E_\uparrow(\mathbf{k}) = E_\downarrow(M_z\mathbf{k}) = E_\downarrow(C_{2z}\mathbf{k}),
  \label{eq:spin-swap}
\end{equation}
to within $11$~meV, where $M_z$ and $C_{2z}$ are the glide and two-fold operations of the $P2_1/c$ space group, respectively. In spin-group notation $[C_2\,\|\,M_z\textit{\textbf{t}}]$ and $[C_2\,\|\,C_{2z}\textit{\textbf{t}}]$ are symmetries of the collinear structure. Because spatial inversion is a symmetry of each spin channel separately, the spin splitting is even ($\Delta E(\mathbf{k}) = \Delta E(-\mathbf{k})$), and it vanishes at every time reversal invariant momentum (TRIM). 
From Eq.~\ref{eq:spin-swap}, one can notice that the spin splitting is odd under both $M_z$ and $C_{2z}$, so it must vanish on the $M_z$ invariant plane $(k_z = 0)$ and along the $C_{2z}$ invariant line ($k_x = k_y = 0$). In Figs.~\ref{fig:altermag_charac}(a) and (b), we show the calculated Wannier-interpolated splitting for bands $B1$ and $B3$, where one can observe that $\Delta E$ vanishes throughout the $k_z=0$ plane and along the $k_x = k_y = 0$ nodal line. 

From our Wannier-interpolated splitting we observe that $\Delta E$ vanishes ($\lesssim 11$~meV) throughout the $k_z=0$ plane and along the $k_x$ and $k_z$ axes, whereas it is finite and sign-changing away from these directions. The even parity, the sign alternation between lobes (shown in Fig.~\ref{fig:altermag_charac} for bands $B1$ and $B3$), and the nodal planes and lines are characteristic of a $d$-wave type form factor. However, due to the monoclinic symmetry of the $P2_1/c$ space group, the spin splitting manifests as a mixed state composed of $d_{xz}$ and $d_{yz}$ components, such that $\Delta E(\textbf{k}) \propto \alpha k_xk_z + \beta  k_yk_z$. The derivation of this symmetry allowed mixed factor via Taylor expansion is provided in Supplementary material Sec. VII.

Within the analyzed Wannier band manifold, the maximum splitting $\Delta E_{\max}(\mathbf{k})=\max_{n\in\mathcal{W}}|E_{n\uparrow}(\mathbf{k})-E_{n\downarrow}(\mathbf{k})|$ reaches approximately 0.19 eV. The largest splittings occur in the occupied bands between $-3$ and $-4$ eV below the Fermi energy, with substantial O-$2p$ and transition metal $e_g$ character. This value should be distinguished from the $\approx 118$ meV splitting reported for the valence band $B1$ along the $T-\Gamma$ path. 
Our $pd$ Hamiltonian also allows us to identify some other important microscopic ingredients responsible for the spin-splittings. The leading nearest neighbor $t_{\text{V-O}}$ and $t_{\text{Ni-O}}$ hopping amplitudes are around 1.9 and 1.35 eV, whereas the direct $t_{\text{V-V}}$ and $t_{\text{Ni-Ni}}$ hoppings are considerably smaller, \textit{viz.} $< 0.17$ eV. 
The inter-sublattice coupling is therefore not direct, but is due to oxygen mediated V-O-Ni superexchange. These TM-ligand hoppings are strongly interorbital and spatially anisotropic in our case. 

To quantify where the observed splittings reside, we calculate the orbital weighted spin-splittings using our $pd$ Hamiltonian. In particular, for each $\mathbf{k}$ we diagonalize $H_{\uparrow}$ and $H_{\downarrow}$ and pair the spin-up and spin-down bands by maximal Wannier-eigenvector overlap, defining $\Delta E_{n}(\mathbf{k}) = E^{\uparrow}_{n\mathbf{k}} - E^{\downarrow}_{n\mathbf{k}}$. For each state, we assign an orbital weight of each group of orbitals $g$ to the splitting, given by
\begin{equation}
w^{g}_{n\mathbf{k}} = \frac{1}{2}\Big[ \sum_{\mu \in g} |\langle \mu|\psi^{\uparrow}_{n\mathbf{k}}\rangle|^{2} + |\langle \mu|\psi^{\downarrow}_{n\mathbf{k}}\rangle|^{2} \Big],
\end{equation}
where $|\mu \rangle$ are the Wannier Functions, and $|\psi^{\sigma}_{n\mathbf{k}}\rangle$ are the eigenstates of $H_{\sigma}(\mathbf{k})$. Thus, we define the fractional contribution of $g$ to the spin-splittings as
\begin{equation}
f_{g} = \frac{\sum_{\mathbf{k}n} |\Delta E_{n}|w^{g}_{n\mathbf{k}}}{\sum_{\mathbf{k}n} |\Delta E_{n}|}.
\end{equation}
We evaluated the values of $f_g$ for bands within the frozen window in a grid of 2500 $\mathbf{k}$ points. For the (LVO)$_1$/(LNO)$_1$ SL, we find that $27\%$, $28\%$, and $45\%$ of the spin splitting come from the V-$d$, Ni-$d$, and O-$p$ Wannier orbitals, respectively. 
Nearly half of the splitting weight thus resides on the oxygen states, showing that the altermagnetism in (LVO)$_1$/(LNO)$_1$ is not a purely $d-d$ phenomenon but a ligand-assisted one. Camerano et al. recently reported a related ligand-mediated mechanism for the $g$-wave altermagnet Co$_{1/4}$NbSe$_2$~\cite{camerano2026ligand}. In our case, the ligand contribution does not originate from ligand holes, which are quenched by the interfacial charge transfer (see Sec.~\ref{subsec:chargetransfer}). Instead, it originates from the strong $p$-$d$ covalency that survives in the charge-transferred state. The leading $t_{\text{V-O}}$ and $t_{\text{Ni-O}}$ hoppings (1.9 and 1.35 eV) exceed the direct $t_{\text{V-V}}$ and $t_{\text{Ni-Ni}}$ hoppings by an order of magnitude, so the eigenstates near the splitting maxima are strongly mixed V t$_{2g}$/Ni $e_g$ - O-$2p$ states. The O-$2p$ orbitals connecting the two antiparallel sublattices inherit their spin polarization and the anisotropy imposed by the tilted octahedra. The O-$2p$ states therefore actively co-carry the momentum dependent spin splitting through covalent mixing with the magnetic $d$ orbitals, rather than merely mediating the superexchange between them.

Further, within each transition metal, the splitting is carried by different groups of orbitals. For vanadium the contribution comes from the $t_{2g}$ manifold, which is dominated by the $d_{xy}$ Wannier Function, in agreement with the V$^{4+}$ 3d$^1$ like picture. On the other hand, Ni contribution comes from both $e_{g}$ orbitals. This orbital selectivity indicates that the microscopic origin of this mixed $d$-wave form factor is the anisotropic, inter orbital V-O-Ni hoppings, which in turn, generate the mixed momentum dependence of the spin splitting.
These hoppings are those connecting the occupied V t$_{2g}$ and the Ni $e_{g}$ states through the tilted oxygen octahedra. It is worth mentioning that this picture can be tested by X-ray spectroscopies (such as XMCD~\cite{xmcd_alter}) at V and Ni $L$ edges, where the ligand-assisted altermagnetism can be tested in (LVO)/(LNO) SLs.  We emphasize that the weights above quantify where the spin splitting concentrates in orbital space, and should not be interpreted as the distribution of the magnetic moments.

\section{\label{sec:conclusions}CONCLUSIONS}

In summary, we have shown that interfacial charge transfer in (LaVO$_3$)$_n$/(LaNiO$_3$)$_m$ superlattices turns the paramagnetic metallic LaNiO$_3$ layers into a magnetically active Ni$^{2+}$ (3d$^8$) sublattice that participates, together with the vanadate layers, in an altermagnetic emergent phase. In the ultrathin limit [(LVO)$_1$/(LNO)$_1$], the resulting spin-splitting of $\approx 118$ meV is about twice that of bulk LaVO$_3$, and it decreases as the nickelate layers become thicker and recover their bulk-like character. According to our calculations, the mechanism is a cooperative interplay between charge transfer and symmetry. The charge transfer drives the system into a $C$-type AFM order, stabilizing a V$^{4+}$ (3d$^1$) and Ni$^{2+}$ (3d$^8$) like configurations. The resulting AFM order, combined with the cooperative octahedral tilts, connects the opposite spin sublattices exclusively through nonsymmorphic screw and glide operations, enforcing a $d$-wave splitting of up to 117.9 meV in the absence of the spin-orbit coupling within a DFT+U insulating state. Orbital-resolved band structures demonstrate that both the vanadium and nickel sublattices actively host the spin splitting, establishing the observed altermagnetism as a collective, charge transfer driven phase rather than a passive proximity effect. 
Although bulk LaNiO$_3$ lies in the negative charge-transfer regime (3$d^8\underline{L}$), the interfacial charge transfer fills the ligand holes. In the (LVO)$_1$/(LNO)$_1$ superlattice, the O-$2p$ states nonetheless carry nearly half of the spin splitting through strong $p-d$ covalency with the magnetic V and Ni sublattices. Dynamic correlations beyond static DFT+U may renormalize the magnitude of the splitting and the size of the local moments, particularly on the vanadium sites. An exemplary case is RuO$_2$, whose altermagnetic order, predicted within DFT based approaches~\cite{Smejkal2020}, is fragile with respect to the value of $U$~\cite{Smolyanyuk2024} and is not observed in muon spin experiments~\cite{Hiraishi2024}. Recent DFT+DMFT calculations show that bulk RuO$_2$ is an itinerant paramagnet located close to the paramagnetic-altermagnetic phase boundary~\cite{Park2026}. In contrast, the present superlattices host well-defined Ni$^{2+}$ (3d$^8$) local moments, and  the $d$-wave nodal structure of the splitting is fixed by the nonsymmorphic spin-group symmetries, which are insensitive to such effects. A finite temperature DFT+DMFT study of our SLs and these effects will be presented elsewhere.

Finally, our results identify interfacial charge transfer in correlated oxide heterostructures as a design principle (tuning knob) for engineering and enhancing altermagnetism, by activating otherwise nonmagnetic constituents such as LaNiO$_3$. We stress that this route is complementary to the search for intrinsic bulk candidates.

\begin{acknowledgments}
The authors acknowledge the financial support from the Air Force Office of Scientific Research (grant number FA9550-24-1-0279). W. H. B. also acknowledges the financial support from the Brazilian agency CNPq (Grants 201149/2024-9, and 305726/2023-4).  W.~H.~B. thank M.~Radovic, T. Schmitt, and C. Piamonteze for useful discussions.
The authors acknowledge the National Laboratory for Scientific Computing (LNCC/MCTI, Brazil) for providing HPC resources of the SDumont supercomputer (URL: http://sdumont.lncc.br), which have contributed to the research results reported within this paper.

\end{acknowledgments}

\appendix

\section{Brillouin Zones for $P2_1/c$ and $P_c$ structures}
\label{appendixA}

\begin{figure}[!htb]
    \centering
    \includegraphics[scale=0.75]{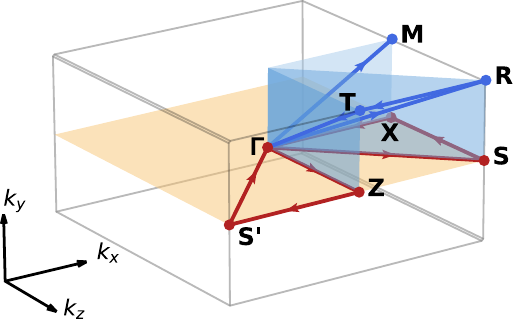}
    \caption{\textbf{Brillouin zone of the $P2_1/c$ space group.} Schematic representation of the bulk brillouin zone for the primitive monoclinic lattice, showing the high-symmetry points and the paths used for the electronic band structure calculations.}
    \label{fig:p21_a}
\end{figure}

\begin{figure}[!htb]
    \centering
    \includegraphics[scale=0.75]{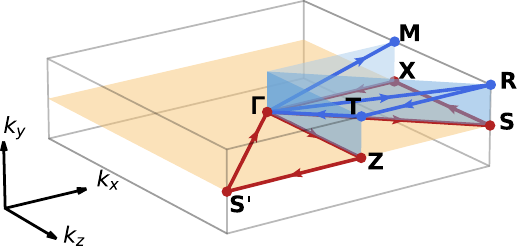}
    \caption{\textbf{Brillouin zone of the $Pc$ space group.} Schematic representation of the bulk brillouin zone for the primitive monoclinic lattice, showing the high-symmetry points and the paths used for the electronic band structure calculations.}
    \label{fig:pc}
\end{figure}

The Brillouin zones and high-symmetry directions considered in our band structure calculations are shown in Figs.~\ref{fig:p21_a} and ~\ref{fig:pc}. The coordinates of high-symmetry $\mathbf{k}$ points mapped in our paths are listed in Table~\ref{tab:high_symmetry_points}.

\begin{table}[htbp]
    \centering
    \renewcommand{\arraystretch}{1.2}
    \caption{\textbf{Coordinates of the high-symmetry $\mathbf{k}$ points.} Fractional coordinates in the Brillouin zone of the primitive monoclinic lattices ($P2_1/c$ and $P_c$) used for the electronic band structure calculations. The coordinates are given in terms of the reciprocal lattice basis vectors.}
    \setlength{\tabcolsep}{4pt}
        \begin{tabular}{@{}cccc@{}}
            \toprule
            High-symmetry point & $k_x$ &$k_y$ &$k_z$  \\
            \midrule
                $\Gamma$& 0.0 & 0.0&0.0\\
                $Z$& 0.0 & 0.0 & 0.5\\
                $S'$& -0.5 & 0.0 & 0.5\\
                $\Gamma$& 0.0 & 0.0&0.0\\
                $S$& 0.5 & 0.0 & 0.5\\
                $X$& 0.5 & 0.0 &0.0\\
                $\Gamma$& 0.0 & 0.0&0.0\\
                $R$&0.5 & 0.5&0.5\\
                $T$&0.0 & 0.5&0.5\\
                $\Gamma$& 0.0 & 0.0&0.0\\
                $M$& 0.5&0.5 &0.0\\
            \bottomrule
        \end{tabular}
    \label{tab:high_symmetry_points}
\end{table}

\bibliography{main}

@article{kohn1965self,
  title={Self-consistent equations including exchange and correlation effects},
  author={Kohn, Walter and Sham, Lu Jeu},
  journal={Physical review},
  volume={140},
  number={4A},
  pages={A1133},
  year={1965},
  publisher={APS},
  doi={10.1103/PhysRev.140.A1133}
}

@article{blochl1994projector,
  author={Bl{\"o}chl, Peter E},
  title={Projector augmented-wave method},
  journal={Physical review B},
  volume={50},
  number={24},
  pages={17953},
  year={1994},
  publisher={APS},
  doi={10.1103/PhysRevB.50.17953}
}

@article{spglib,
  author = {Atsushi Togo and Kohei Shinohara and Isao Tanaka},
  title = {Spglib: a software library for crystal symmetry search},
  journal = {Sci. Technol. Adv. Mater., Meth.},
  volume = {4},
  number = {1},
  pages = {2384822--2384836},
  year = {2024},
  doi = {10.1080/27660400.2024.2384822},
}

@article{lnobulk,
  title = {Lattice normal modes and electronic properties of the correlated metal {LaNiO$_{3}$}},
  author = {Gou, Gaoyang and Grinberg, Ilya and Rappe, Andrew M. and Rondinelli, James M.},
  journal = {Phys. Rev. B},
  volume = {84},
  issue = {14},
  pages = {144101},
  numpages = {13},
  year = {2011},
  month = {Oct},
  publisher = {American Physical Society},
  doi = {10.1103/PhysRevB.84.144101},
}

@article{lvobulk,
title = {Crystal structure of {LaVO$_3$} by Rietveld refinement method},
journal = {Materials Letters},
volume = {58},
number = {11},
pages = {1737-1740},
year = {2004},
issn = {0167-577X},
doi = {https://doi.org/10.1016/j.matlet.2003.10.059},
author = {R.T.A. Khan and J. Bashir and N. Iqbal and M.Nasir Khan}
}

@article{monkhorst1976special,
  title={Special points for Brillouin-zone integrations},
  author={Monkhorst, Hendrik J and Pack, James D},
  journal={Physical review B},
  volume={13},
  number={12},
  pages={5188},
  year={1976},
  publisher={APS},
  doi={10.1103/PhysRevB.13.5188}
}

@article{liechtenstein1995density,
  title={Density-functional theory and strong interactions: Orbital ordering in Mott-Hubbard insulators},
  author={Liechtenstein, AI and Anisimov, Vladimir I and Zaanen, Jan},
  journal={Physical Review B},
  volume={52},
  number={8},
  pages={R5467},
  year={1995},
  publisher={APS},
  doi={10.1103/PhysRevB.52.R5467}
}

@article{vsmejkal2022beyond,
  title={Beyond conventional ferromagnetism and antiferromagnetism: A phase with nonrelativistic spin and crystal rotation symmetry},
  author={{\v{S}}mejkal, Libor and Sinova, Jairo and Jungwirth, Tomas},
  journal={Physical Review X},
  volume={12},
  number={3},
  pages={031042},
  year={2022},
  publisher={APS},
  doi={10.1103/PhysRevX.12.031042}
}

@article{vsmejkal2022emerging,
  title={Emerging research landscape of altermagnetism},
  author={{\v{S}}mejkal, Libor and Sinova, Jairo and Jungwirth, Tomas},
  journal={Physical Review X},
  volume={12},
  number={4},
  pages={040501},
  year={2022},
  publisher={APS},
  doi={10.1103/PhysRevX.12.040501}
}

@article{mazin_altermag2022,
  title = {Editorial: Altermagnetism---A New Punch Line of Fundamental Magnetism},
  author = {Mazin, Igor},
  collaboration = {The PRX Editors},
  journal = {Phys. Rev. X},
  volume = {12},
  issue = {4},
  pages = {040002},
  numpages = {3},
  year = {2022},
  month = {Dec},
  publisher = {American Physical Society},
  doi = {10.1103/PhysRevX.12.040002},
}

@Article{Jungwirth2026,
author={Jungwirth, T.
and Sinova, J.
and Wadley, P.
and Kriegner, D.
and Reichlov{\'a}, H.
and Krizek, F.
and Ohno, H.
and {\v{S}}mejkal, L.},
title={Altermagnetic spintronics},
journal={Nature Physics},
year={2026},
month={Jul},
day={01},
volume={22},
number={7},
pages={1012-1021},
issn={1745-2481},
doi={10.1038/s41567-026-03337-w},
}

@article{roigPRBMinimalModel,
  title = {Minimal models for altermagnetism},
  author = {Roig, Merc\`e and Kreisel, Andreas and Yu, Yue and Andersen, Brian M. and Agterberg, Daniel F.},
  journal = {Phys. Rev. B},
  volume = {110},
  issue = {14},
  pages = {144412},
  numpages = {20},
  year = {2024},
  month = {Oct},
  publisher = {American Physical Society},
  doi = {10.1103/PhysRevB.110.144412},
}

@article{Bai_AdvFunc_2026,
author = {Bai, Ling and Feng, Wanxiang and Liu, Siyuan and Šmejkal, Libor and Mokrousov, Yuriy and Yao, Yugui},
title = {Altermagnetism: Exploring New Frontiers in Magnetism and Spintronics},
journal = {Advanced Functional Materials},
volume = {34},
number = {49},
pages = {2409327},
doi = {https://doi.org/10.1002/adfm.202409327},
year = {2024}
}

@article{Cui_2023PRB,
  title = {Giant spin-Hall and tunneling magnetoresistance effects based on a two-dimensional nonrelativistic antiferromagnetic metal},
  author = {Cui, Qirui and Zhu, Yingmei and Yao, Xiong and Cui, Ping and Yang, Hongxin},
  journal = {Phys. Rev. B},
  volume = {108},
  issue = {2},
  pages = {024410},
  numpages = {9},
  year = {2023},
  month = {Jul},
  publisher = {American Physical Society},
  doi = {10.1103/PhysRevB.108.024410},
}

@article{magnon_topologyPRB,
  title = {Magnon topology driven by altermagnetism},
  author = {Khatua, Subhankar and Kravchuk, Volodymyr P. and Yershov, Kostiantyn V. and van den Brink, Jeroen},
  journal = {Phys. Rev. B},
  volume = {112},
  issue = {21},
  pages = {214422},
  numpages = {16},
  year = {2025},
  month = {Dec},
  publisher = {American Physical Society},
  doi = {10.1103/xg1x-sj4c},
}

@Article{Guerrero-Sanchez2026,
author={Guerrero-Sanchez, J.
and Ponce-Perez, R.
and Hoat, D. M.
and Gonzalez-Hernandez, R.},
title={Emergent altermagnetism and topological response in Janus MnPSX monolayers},
journal={Scientific Reports},
year={2026},
month={Mar},
day={11},
volume={16},
number={1},
pages={13056},
issn={2045-2322},
doi={10.1038/s41598-026-38927-4},
}

@article{spaldin1_PRX2024,
  title = {Ferroically Ordered Magnetic Octupoles in $d$-Wave Altermagnets},
  author = {Bhowal, Sayantika and Spaldin, Nicola A.},
  journal = {Phys. Rev. X},
  volume = {14},
  issue = {1},
  pages = {011019},
  numpages = {19},
  year = {2024},
  month = {Feb},
  publisher = {American Physical Society},
  doi = {10.1103/PhysRevX.14.011019},
}

@article{picozzi_PRB2025,
  title = {Designing nonrelativistic spin splitting in oxide perovskites},
  author = {Bandyopadhyay, Subhadeep and Picozzi, Silvia and Bhowal, Sayantika},
  journal = {Phys. Rev. B},
  volume = {112},
  issue = {6},
  pages = {064405},
  numpages = {15},
  year = {2025},
  month = {Aug},
  publisher = {American Physical Society},
  doi = {10.1103/1r6k-s46h},
}

@article{Cano_2025JAP,
    author = {Bernardini, Fabio and Fiebig, Manfred and Cano, Andrés},
    title = {Ruddlesden–Popper and perovskite phases as a material platform for altermagnetism},
    journal = {Journal of Applied Physics},
    volume = {137},
    number = {10},
    pages = {103903},
    year = {2025},
    month = {03},
    issn = {0021-8979},
    doi = {10.1063/5.0252836},
}

@article{Ganguli_PRB2025,
  title = {Altermagnetism in the orthorhombic {$Pnma$} structure through group theory and DFT calculations},
  author = {Rooj, Suman and Saxena, Sugandha and Ganguli, Nirmal},
  journal = {Phys. Rev. B},
  volume = {111},
  issue = {1},
  pages = {014434},
  numpages = {11},
  year = {2025},
  month = {Jan},
  publisher = {American Physical Society},
  doi = {10.1103/PhysRevB.111.014434},
}

@article{Chakhalian_RMP2014,
  title = {Colloquium: Emergent properties in plane view: Strong correlations at oxide interfaces},
  author = {Chakhalian, Jak and Freeland, John W. and Millis, Andrew J. and Panagopoulos, Christos and Rondinelli, James M.},
  journal = {Rev. Mod. Phys.},
  volume = {86},
  issue = {4},
  pages = {1189--1202},
  numpages = {14},
  year = {2014},
  month = {Oct},
  publisher = {American Physical Society},
  doi = {10.1103/RevModPhys.86.1189},
}

@article{DFT1,
  title = {Inhomogeneous Electron Gas},
  author = {Hohenberg, P. and Kohn, W.},
  journal = {Phys. Rev.},
  volume = {136},
  issue = {3B},
  pages = {B864--B871},
  numpages = {0},
  year = {1964},
  month = {Nov},
  publisher = {American Physical Society},
  doi = {10.1103/PhysRev.136.B864},
}

@article{pbesol_perdew,
  title = {Restoring the Density-Gradient Expansion for Exchange in Solids and Surfaces},
  author = {Perdew, John P. and Ruzsinszky, Adrienn and Csonka, G\'abor I. and Vydrov, Oleg A. and Scuseria, Gustavo E. and Constantin, Lucian A. and Zhou, Xiaolan and Burke, Kieron},
  journal = {Phys. Rev. Lett.},
  volume = {100},
  issue = {13},
  pages = {136406},
  numpages = {4},
  year = {2008},
  month = {Apr},
  publisher = {American Physical Society},
  doi = {10.1103/PhysRevLett.100.136406},
}

@article{vasp1,
title = {Efficiency of ab-initio total energy calculations for metals and semiconductors using a plane-wave basis set},
author = "G. Kresse and J. Furthm$\rm\ddot{u}$ller",
journal = "Comput. Mater. Sci.",
volume = "6", 
pages = "15",
year = "1996",
doi={10.1016/0927-0256(96)00008-0}
}

@article{vasp2,
 title = {Efficient iterative schemes for ab initio total-energy calculations using a plane-wave basis set},
author = "G. Kresse and J. Furthm$\rm\ddot{u}$ller",
journal = "Phys. Rev. B",
volume = "54", 
pages = "11169",
year = "1996",
doi={10.1103/PhysRevB.54.11169}
}

@article{MLWFs_marzari,
  title = {Maximally localized Wannier functions: Theory and applications},
  author = {Marzari, Nicola and Mostofi, Arash A. and Yates, Jonathan R. and Souza, Ivo and Vanderbilt, David},
  journal = {Rev. Mod. Phys.},
  volume = {84},
  issue = {4},
  pages = {1419--1475},
  numpages = {0},
  year = {2012},
  month = {Oct},
  publisher = {American Physical Society},
  doi = {10.1103/RevModPhys.84.1419},
}

@article{wannier90,
	year = 2020,
	month = {jan},
	publisher = {{IOP} Publishing},
	volume = {32},
	number = {16},
	pages = {165902},
	author = {Giovanni Pizzi and Valerio Vitale and Ryotaro Arita and Stefan Blügel and Frank Freimuth and Guillaume G{\'{e}}ranton and Marco Gibertini and Dominik Gresch and Charles Johnson and Takashi Koretsune and Julen Iba{\~{n}}ez-Azpiroz and Hyungjun Lee and Jae-Mo Lihm and Daniel Marchand and Antimo Marrazzo and Yuriy Mokrousov and Jamal I Mustafa and Yoshiro Nohara and Yusuke Nomura and Lorenzo Paulatto and Samuel Ponc{\'{e}} and Thomas Ponweiser and Junfeng Qiao and Florian Thöle and Stepan S Tsirkin and Ma{\l}gorzata Wierzbowska and Nicola Marzari and David Vanderbilt and Ivo Souza and Arash A Mostofi and Jonathan R Yates},
	title = {Wannier90 as a community code: new features and applications},
	journal = {Journal of Physics: Condensed Matter},
    doi={10.1088/1361-648X/ab51ff}
}

@article{Vanad_PRBUeff1,
  title = {Layer-dependent electronic structures and magnetic ground states of polar-polar {LaVO$_3$/KTaO$_3$} (001) heterostructures},
  author = {Patel, Shubham and Mohanta, Narayan and Nandy, Snehasish and Mahanti, Subhendra D. and Taraphder, A.},
  journal = {Phys. Rev. B},
  volume = {110},
  issue = {5},
  pages = {054402},
  numpages = {13},
  year = {2024},
  month = {Aug},
  publisher = {American Physical Society},
  doi = {10.1103/PhysRevB.110.054402},
}

@article{Ni_PRBUeff1,
  title = {Confinement-induced metal-to-insulator transition in strained {LaNiO$_{3}$/LaAlO$_{3}$} superlattices},
  author = {Blanca-Romero, Ariadna and Pentcheva, Rossitza},
  journal = {Phys. Rev. B},
  volume = {84},
  issue = {19},
  pages = {195450},
  numpages = {7},
  year = {2011},
  month = {Nov},
  publisher = {American Physical Society},
  doi = {10.1103/PhysRevB.84.195450},
}

@article{GKA,
title = {Superexchange interaction and symmetry properties of electron orbitals},
journal = {Journal of Physics and Chemistry of Solids},
volume = {10},
number = {2},
pages = {87-98},
year = {1959},
issn = {0022-3697},
doi = {https://doi.org/10.1016/0022-3697(59)90061-7},
author = {Junjiro Kanamori}
}

@article{Smejkal2020,
  author  = {{\v{S}}mejkal, Libor and Gonz{\'a}lez-Hern{\'a}ndez, Rafael and Jungwirth, T. and Sinova, J.},
  title   = {Crystal time-reversal symmetry breaking and spontaneous {Hall} effect in collinear antiferromagnets},
  journal = {Sci. Adv.},
  volume  = {6},
  pages   = {eaaz8809},
  year    = {2020},
  doi     = {10.1126/sciadv.aaz8809}
}

@article{Hiraishi2024,
  author  = {Hiraishi, M. and Okabe, H. and Koda, A. and Kadono, R. and Muroi, T. and Hirai, D. and Hiroi, Z.},
  title   = {Nonmagnetic Ground State in {RuO$_2$} Revealed by Muon Spin Rotation},
  journal = {Phys. Rev. Lett.},
  volume  = {132},
  pages   = {166702},
  year    = {2024},
  doi     = {10.1103/PhysRevLett.132.166702}
}

@article{Smolyanyuk2024,
  author  = {Smolyanyuk, Andriy and Mazin, Igor I. and Garcia-Gassull, Laura and Valent{\'\i}, Roser},
  title   = {Fragility of the magnetic order in the prototypical altermagnet {RuO$_2$}},
  journal = {Phys. Rev. B},
  volume  = {109},
  pages   = {134424},
  year    = {2024},
  doi     = {10.1103/PhysRevB.109.134424}
}

@misc{Park2026,
  author  = {Park, Ina and Kim, Dongwook and Lee, Inho and Hong, Jisook and Goh, Beomjoon and Jang, Bo Gyu},
  title   = {Correlation-driven tunability of altermagnetism in {RuO$_2$}},
  eprint={2605.13559},
  archivePrefix={arXiv},
  primaryClass={cond-mat.mtrl-sci},
  year    = {2026},
  url={10.48550/arXiv.2605.13559}
}

@misc{camerano2026ligand,
      title={Ligand-mediated Origin of Altermagnetic Spin-Splitting}, 
      author={Luigi Camerano and Federico Bisti and Gianni Profeta},
      year={2026},
      eprint={2605.21284},
      archivePrefix={arXiv},
      primaryClass={cond-mat.mtrl-sci},
      url={https://arxiv.org/abs/2605.21284}, 
}

@article{LVO_exp1,
  title = {Orbital-ordering-induced phase transition in {LaVO$_{3}$} and {CeVO$_{3}$}},
  author = {Ren, Y. and Nugroho, A. A. and Menovsky, A. A. and Strempfer, J. and R\"utt, U. and Iga, F. and Takabatake, T. and Kimball, C. W.},
  journal = {Phys. Rev. B},
  volume = {67},
  issue = {1},
  pages = {014107},
  numpages = {6},
  year = {2003},
  month = {Jan},
  publisher = {American Physical Society},
  doi = {10.1103/PhysRevB.67.014107},
  url = {https://link.aps.org/doi/10.1103/PhysRevB.67.014107}
}

@article{LVO_exp2,
  title = {Spin, orbital ordering, and magnetic dynamics of {LaVO$_{3}$}: Magnetization, heat capacity, and neutron scattering studies},
  author = {Tung, L. D. and Ivanov, A. and Schefer, J. and Lees, M. R. and Balakrishnan, G. and Paul, D. McK.},
  journal = {Phys. Rev. B},
  volume = {78},
  issue = {5},
  pages = {054416},
  numpages = {5},
  year = {2008},
  month = {Aug},
  publisher = {American Physical Society},
  doi = {10.1103/PhysRevB.78.054416},
  url = {https://link.aps.org/doi/10.1103/PhysRevB.78.054416}
}

@article{imada1998metal,
  title={Metal-insulator transitions},
  author={Imada, Masatoshi and Fujimori, Atsushi and Tokura, Yoshinori},
  journal={Reviews of modern physics},
  volume={70},
  number={4},
  pages={1039},
  year={1998},
  publisher={APS},
  doi={10.1103/RevModPhys.70.1039}
}

@article{Tokura_Science2000,
author = {Y. Tokura  and N. Nagaosa },
title = {Orbital Physics in Transition-Metal Oxides},
journal = {Science},
volume = {288},
number = {5465},
pages = {462-468},
year = {2000},
doi = {10.1126/science.288.5465.462}}

@Article{Hwang2012,
author={Hwang, H. Y.
and Iwasa, Y.
and Kawasaki, M.
and Keimer, B.
and Nagaosa, N.
and Tokura, Y.},
title={Emergent phenomena at oxide interfaces},
journal={Nature Materials},
year={2012},
month={Feb},
day={01},
volume={11},
number={2},
pages={103-113},
issn={1476-4660},
doi={10.1038/nmat3223},
url={https://doi.org/10.1038/nmat3223}
}

@article{naka2025altermagnetic,
  title={Altermagnetic perovskites},
  author={Naka, Makoto and Motome, Yukitoshi and Seo, Hitoshi},
  journal={npj Spintronics},
  volume={3},
  number={1},
  pages={1},
  year={2025},
  publisher={Nature Publishing Group UK London},
  doi={10.1038/s44306-024-00066-9}
}

@article{catalano2018rare,
  title={Rare-earth nickelates \ce{RNiO3}: thin films and heterostructures},
  author={Catalano, Sara and Gibert, M and Fowlie, J and {\'I}{\~n}iguez, J and Triscone, JM and Kreisel, Jens},
  journal={Reports on Progress in Physics},
  volume={81},
  number={4},
  pages={046501},
  year={2018},
  publisher={IOP Publishing},
  doi={10.1088/1361-6633/aaa37a}
}

@Article{xmcd_alter,
author={Hariki, A.
and Sakurai, K.
and Okauchi, T.
and Kune{\v{s}}, J.},
title={Separating altermagnetic and ferromagnetic effects in {X}-ray magnetic dichroism of rutile \ce{NiF2}},
journal={npj Quantum Materials},
year={2025},
month={May},
day={17},
volume={10},
number={1},
pages={49},
issn={2397-4648},
doi={10.1038/s41535-025-00753-8},
url={https://doi.org/10.1038/s41535-025-00753-8}
}

@article{boris2011dimensionality,
  title={Dimensionality control of electronic phase transitions in nickel-oxide superlattices},
  author={Boris, AV and Matiks, Y and Benckiser, E and Frano, A and Popovich, P and Hinkov, V and Wochner, P and Castro-Colin, M and Detemple, E and Malik, Vivek Kumar and others},
  journal={Science},
  volume={332},
  number={6032},
  pages={937--940},
  year={2011},
  publisher={American Association for the Advancement of Science},
  doi={10.1126/science.1202647}
}

@article{ishizaka2011giant,
  title={Giant Rashba-type spin splitting in bulk {BiTeI}},
  author={Ishizaka, K and Bahramy, Mohammad Saeed and Murakawa, H and Sakano, M and Shimojima, T and Sonobe, T and Koizumi, K and Shin, S and Miyahara, H and Kimura, A and others},
  journal={Nature materials},
  volume={10},
  number={7},
  pages={521--526},
  year={2011},
  publisher={Nature Publishing Group UK London},
  doi={10.1038/nmat3051}
}

@article{beck2019charge,
  title={Charge transfer in {{LaVO$_3$}/{LaTiO$_3$}} multilayers: strain-controlled dimensionality of interface metallicity between two Mott insulators},
  author={Beck, Sophie and Ederer, Claude},
  journal={arXiv preprint arXiv:1905.00290},
  year={2019}
}

@article{beck2023interfacial,
  title={Interfacial doping in {LaVO$_3$}/{SrVO$_3$} multilayers from {DFT+DMFT}},
  author={Beck, Sophie and Ederer, Claude},
  journal={Physical Review Materials},
  volume={7},
  number={5},
  pages={055003},
  year={2023},
  publisher={APS},
  doi={10.1103/PhysRevMaterials.7.055003}
}

@article{asmara2024emergence,
  title={Emergence of Interfacial Magnetism in Strongly-Correlated Nickelate-Titanate Superlattices},
  author={Asmara, Teguh Citra and Green, Robert J and Suter, Andreas and Wei, Yuan and Zhang, Wenliang and Knez, Daniel and Harris, Grant and Tseng, Yi and Yu, Tianlun and Betto, Davide and others},
  journal={Advanced Materials},
  volume={36},
  number={38},
  pages={2310668},
  year={2024},
  publisher={Wiley Online Library},
  doi={10.1002/adma.202310668}
}

@article{zhong2017band,
  title={Band alignment and charge transfer in complex oxide interfaces},
  author={Zhong, Zhicheng and Hansmann, Philipp},
  journal={Physical Review X},
  volume={7},
  number={1},
  pages={011023},
  year={2017},
  publisher={APS},
  doi={10.1103/PhysRevX.7.011023}
}

@article{mercy2017structurally,
  title={Structurally triggered metal-insulator transition in rare-earth nickelates},
  author={Mercy, Alain and Bieder, Jordan and {\'I}{\~n}iguez, Jorge and Ghosez, Philippe},
  journal={Nature communications},
  volume={8},
  number={1},
  pages={1677},
  year={2017},
  publisher={Nature Publishing Group UK London},
  doi={10.1038/s41467-017-01811-x}
}

@article{varignon2017complete,
  title={Complete phase diagram of rare-earth nickelates from first-principles},
  author={Varignon, Julien and Grisolia, Mathieu N and {\'I}{\~n}iguez, Jorge and Barth{\'e}l{\'e}my, Agn{\`e}s and Bibes, Manuel},
  journal={npj Quantum Materials},
  volume={2},
  number={1},
  pages={21},
  year={2017},
  publisher={Nature Publishing Group UK London},
  doi={10.1038/s41535-017-0024-9}
}

@article{zhu2026altermagnetic,
  title={Altermagnetic proximity effect},
  author={Zhu, Ziye and Huang, Richang and Chen, Xianzhang and Cui, Zhou and Duan, Xunkai and Zhang, Jiayong and {\v{Z}}uti{\'c}, Igor and Zhou, Tong},
  journal={Physical Review Letters},
  volume={136},
  number={18},
  pages={186702},
  year={2026},
  publisher={APS},
  doi={10.1103/kqy8-myz1}
}

@article{arima1993variation,
  title = {Variation of optical gaps in perovskite-type 3d transition-metal oxides},
  author = {Arima, T. and Tokura, Y. and Torrance, J. B.},
  journal = {Phys. Rev. B},
  volume = {48},
  issue = {23},
  pages = {17006--17009},
  numpages = {0},
  year = {1993},
  month = {Dec},
  publisher = {American Physical Society},
  doi = {10.1103/PhysRevB.48.17006},
  url = {https://link.aps.org/doi/10.1103/PhysRevB.48.17006}
}

@article{assmann2013oxide,
  title = {Oxide Heterostructures for Efficient Solar Cells},
  author = {Assmann, Elias and Blaha, Peter and Laskowski, Robert and Held, Karsten and Okamoto, Satoshi and Sangiovanni, Giorgio},
  journal = {Phys. Rev. Lett.},
  volume = {110},
  issue = {7},
  pages = {078701},
  numpages = {5},
  year = {2013},
  month = {Feb},
  publisher = {American Physical Society},
  doi = {10.1103/PhysRevLett.110.078701},
  url = {https://link.aps.org/doi/10.1103/PhysRevLett.110.078701}
}

@article{anas2023structural,
  title={Structural and magnetic properties of {LaVO$_3$}-Absence of anomalous diamagnetism},
  author={Anas, M and Jain, Anil and Gupta, Mukul and Sagdeo, Archna and Yusuf, SM and Maitra, T and Malik, VK},
  journal={Ceramics International},
  volume={49},
  number={6},
  pages={9672--9680},
  year={2023},
  publisher={Elsevier},
  doi={10.1016/j.ceramint.2022.11.138}
}

\end{document}


\title{Supplementary Material ``Emergent altermagnetism in (LaVO$_3$)$_n$/(LaNiO$_3$)$_m$ superlattices from interfacial charge transfer and octahedral tilts"}

\author{Gustavo H. Cassemiro}
\affiliation{Departamento de Física, Universidade Federal de Minas Gerais, 
C. P. 702, 30123-970, Belo Horizonte, MG, Brazil}
\affiliation{Instituto de F\'{i}sica, Universidade de S\~ao Paulo, 05508-090 São Paulo, Brazil}
\author{Walber H. Brito}
\affiliation{Instituto de F\'{i}sica, Universidade de S\~ao Paulo, 05508-090 São Paulo, Brazil}

\maketitle

\section{The U-dependence of collinear magnetic ground state in (LVO)$_1$/(LNO)$_1$}\label{ap:U_dep_superlattices}

\begin{figure}[!htb]
    \centering
    \includegraphics[scale=1]{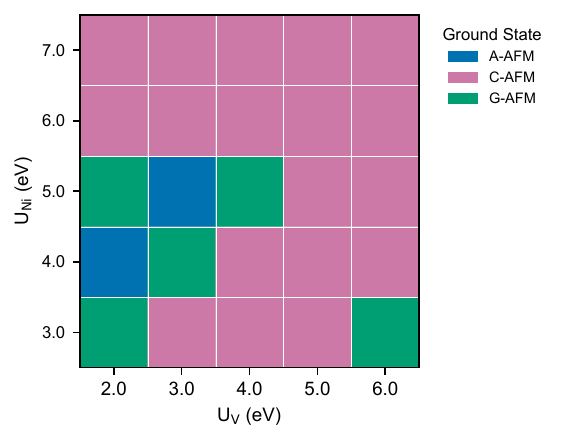}
    \caption{\textbf{Magnetic phase stability diagram of the (LaVO$_3$)$_1$/(LaNiO$_3$)$_1$ superlattice.} The diagram shows the robustness of the $C$-type antiferromagnetic ground state (highlighted in pink) across the Hubbard parameter space.}
    \label{fig:u_dependence}
\end{figure}

To ensure that the stabilization of the $C$-type antiferromagnetic ($C$-AFM) ground state is not an artifact of a specific parameter choice, we evaluated the relative phase stability across a phase space of U. Using the (LaVO$_3$)$_1$/(LaNiO$_3$)$_1$ as the reference system, we systematically varied both U$_V$ and U$_{Ni}$. For each (U$_V$,U$_{Ni}$) pair, the total energies of the collinear magnetic configurations (e.g., $A$-AFM, $C$-AFM, $G$-AFM and FM) was calculated and compared as shown in Fig.~\ref{fig:u_dependence}.

Our results shows that the $C$-AFM phase remains the magnetic ground state across the wide range of evaluated Hubbard parameters, in particular, for sizable U values. This behavior indicates that the energetic preference for the $C$-AFM order is driven by the interfacial charge transfer and the octahedra distortions, rather than being balanced by the magnitude of the Hubbard parameter. 

\section{Magnetic ground state and control calculations}
\label{sec:sm_mag_ground_conf}

To understand in more details the effects of the structural distortion and chemical composition,  on the magnetic ground state of (LNO)$_1$/(LVO)$_1$ SL, we performed two sets of control calculations. 

First, to evaluate whether structural distortions alone are sufficient to stabilize the C-AFM ground state, we performed a calculation in which all V atoms in the (LVO)$_1$/(LNO)$_1$ SL were replaced by Ni, maintaining the $P2_1/c$ symmetry, lattice parameters, and atomic positions of the relaxed SL. As shown in Tabel~\ref{tab:mag_LaNiO3_p21a_scf}, imposing the superlattice geometry on a LaNiO$_3$ system yields a FM ground state, with the $C$-AFM lying $142.1$ (meV/f.u.) higher in energy. This indicates that structural distortions alone cannot account for the observed magnetic order. The chemical composition and the resulting interfacial charge transfer from vanadate to the nicklate layers are essential to stabilize the $C$-AFM state that hosts the altermagnetic phase. 

\begin{table}[htb]
    \centering
    \caption{\textbf{Relative energies $\Delta E$ of collinear magnetic configurations for the constrained LaNiO$_3$ system.} Values of $\Delta E$ are shown for a LaNiO$_3$ lattice constrained to the fully relaxed lattice parameters and atomic positions of the (LVO)$_1$/(LNO)$_1$ superlattice ($P2_1/c$ space group).}
    \renewcommand{\arraystretch}{1.2}
    \setlength{\tabcolsep}{8pt}
    \begin{tabular}{@{}cccc@{}}
        \toprule \hline 
        System & Space group & Magnetic Configuration & $\Delta E$ (meV/f.u.) \\ 
        \midrule
         &  & FM (GS) & 0.0 \\
        LaNiO$_3$ & $P2_1/c$& A-AFM & 42.6\\
         & & C-AFM& 142.1\\
        &  & G-AFM & 339.1\\
        \bottomrule 
    \end{tabular}
    \label{tab:mag_LaNiO3_p21a_scf}
\end{table}

In addition, we build the (LVO)$_1$/(LNO)$_1$ SL to an untilted tetragonal geometry (space group $P4/mmm$). As shown in Table~\ref{tab:mag_LVO_LNO_no_rotations}, for the standard interaction parameters ($U_V=3.8$ eV, $U_{Ni}=4.8$ eV, $J=0.8$ eV), the ferromagnetic (FM) state emerges as the minimum, rather than the $C$-AFM state observed in the fully relaxed structure. Further, only for $J=0$, we obtained AFM configurations.
This behavior is discussed in the main text.

\begin{table}[htbp]
    \centering
    \caption{\textbf{Relative energies of collinear magnetic configurations for te idealized untilted superlattice.} Values of $\Delta E$ are shown for the LVO)$_1$/(LNO)$_1$ structure constrained to the $P4/mmm$ space group, calculated with $J=0.8$ eV and $J=0.0$ eV. For $J=0.8$ eV, all initialized AFM configurations relaxed into Ferrimagnetic (FiM) states due to incomplete moment cancellation. In the $J=0.0$ eV case, the initial $C$-AFM configuration relaxed into the $G$-AFM state.
    }
    \renewcommand{\arraystretch}{1.2}
    \setlength{\tabcolsep}{8pt}
    \begin{tabular}{@{}cccc@{}}
        \toprule \hline 
        System & Space group & Magnetic Configuration & $\Delta E$ (meV/f.u.) \\ 
        \midrule
         &  &  FM (GS) & 0.0 \\
        (LaVO$_3$)$_1$/(LaNiO$_3$)$_1$ & $P4/mmm$& A-AFM $\rightarrow$ FiM & 19.9\\
        ($J=0.8$ eV) &  & C-AFM $\rightarrow$ FiM& 27.3 \\
        &  & G-AFM $\rightarrow$ FiM& 38.7 \\
        \midrule
        &  &  FM (GS) & 0.0 \\
        (LaVO$_3$)$_1$/(LaNiO$_3$)$_1$ & $P4/mmm$& G-AFM & 18.4 \\
        ($J=0.0$ eV) &  & C-AFM $\rightarrow$ G-AFM & 18.4 \\
        &  & A-AFM & 26.4 \\
        \bottomrule 
    \end{tabular}
    \label{tab:mag_LVO_LNO_no_rotations}
\end{table}

Finally, to verify whether the FM ground state in the $P4/mmm$ structure is driven solely by the Hund coupling, we performed additional calculations using $J=0$ eV. As displayed in Table~\ref{tab:mag_LVO_LNO_no_rotations}, suppressing the atomic Hund's coupling alters the stability among the magnetic configurations, although the FM remains the ground state. The $G$-AFM state becomes significantly more stable, and the $C$-AFM configuration collapses into the $G$-AFM state during the self-consistent cycle. 

\begin{figure}[H]
    \centering
    \includegraphics[scale=1]{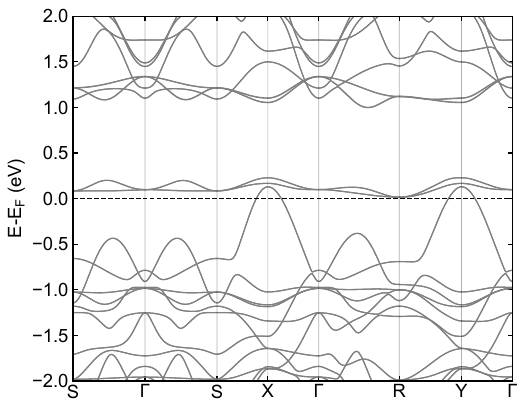}
    \caption{\textbf{Spin-resolved electronic band structure of the untilted (LaVO$_3$)$_1$/(LaNiO$_3$)$_1$ superlattice.} The band dispersion is shown for the idealized $P4/mmm$ space group with $G$-AFM configuration at $J=0$ eV. The absence of the cooperative octahedral tilts restores the translational symmetries connecting the sublattices and completely suppressing the momentum-dependent nonrelativistic spin splitting observed in the distorted $P2_1/c$ phase. 
    }
    \label{fig:tetra_band}
\end{figure}

Therefore, the absence of the octahedral tilts has an important impact on the magnetic and electronic structure. In Fig.~\ref{fig:tetra_band}, we present the band structure for the untilted $P4/mmm$ configuration for the $J=0$ eV case. Unlike the fully relaxed $P2_1/c$ structure, the idealized tetragonal geometry suppresses the altermagnetic $d$-wave spin splitting. Without the GdFeO$_3$-type distortions, the opposite spin sublattices are connected by lattice translations in the AFM metastable states, that generates a global zero spin splitting ($\Delta E(\mathbf{k}) = 0$).

\section{Charge transfer}
\label{sec:sm_charge_transfer}

\begin{figure}[!htb]
    \centering
    \includegraphics[scale=0.8]{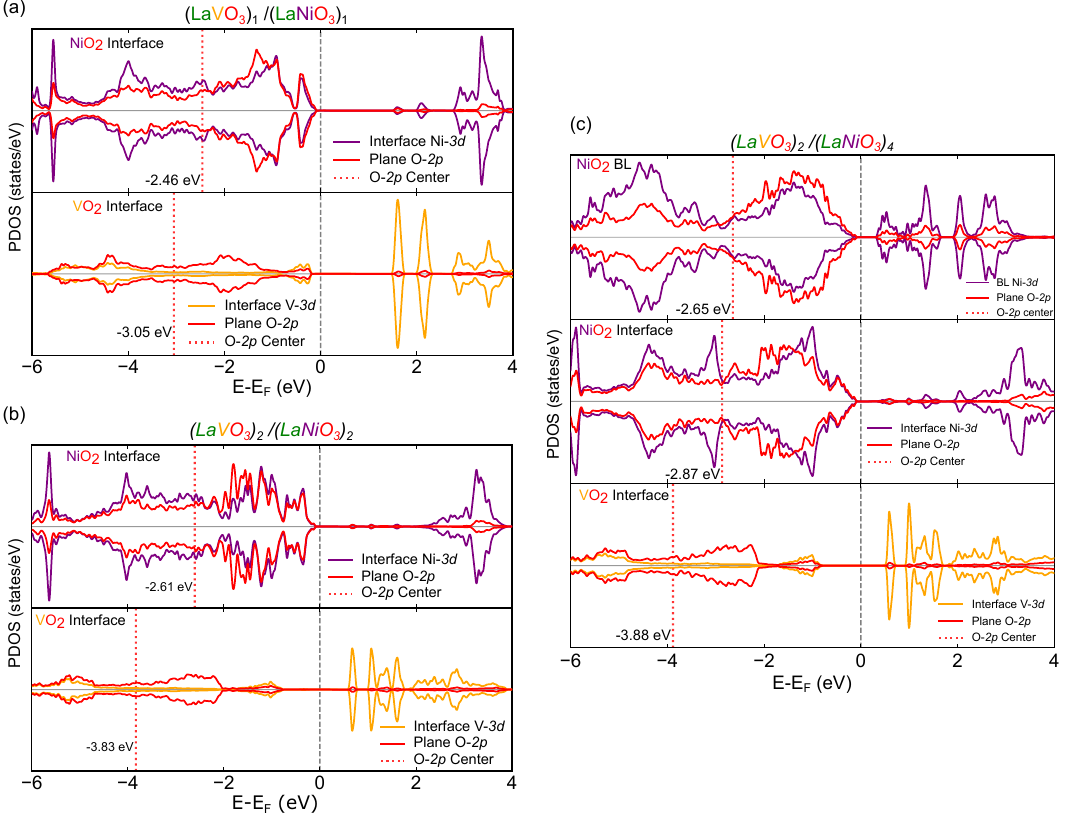}
    \caption{\textbf{Layer-resolved projected density of states (PDOS).} The PDOS is shown for the (a) (LVO)$_1$/(LNO)$_1$, (b) (LVO)$_2$/(LNO)$_2$, and (c) (LVO)$_2$/(LNO)$_4$ superlattices. The dashed red lines show the evolution of the O $2p$ band centers across the interface, highlighting the local electrostatic potential drop that accommodates the charge transfer.}
    \label{fig:pdos_layer_complete}
\end{figure}

The formation of the (LVO)$_n$/(LNO)$_m$ superlattices induces a electronic reconstruction driven by an interfacial charge transfer between the vanadate and nickelate layers. In our work we observed this phenomenon in the layer-resolved projected density of states (PDOS). As shown in Fig.~\ref{fig:pdos_layer_complete}, the PDOS reveal that the V $t_{2g}$ states are depleted pushed above the Fermi Level, while the Ni $e_g$ states become filled. By equalizing the electronic chemical potential across the interface, this charge redistribution effectively quenches the Ni ligant holes and establishes the emergent V$^{4+}$ ($3d^{1}$) and Ni$^{2+}$ ($3d^{8}$) that drive the localized magnetic moments that govern the magnetic order of the SL.

To understand how the system spatially accommodates this charge transfer, we screen the layer dependent band shifts. As can be seen in the PDOS profiles, the centers of the O $2p$ bands exhibit a local electrostatic potential drop across the interface to counterbalance the initial mismatch of the bulk oxygen bands. For example, in ultrathin (LVO)$_1$/(LNO)$_1$ SL, the O $2p$ center shifts from $-3.05$ eV at the VO$_2$ interface to $-2.46$ eV at the NiO$_2$ interface. In the framework of oxide heterojunctions established by Zhong and Hansmann~\cite{zhong2017band}, this spatial modulation screens the built-in potential.

\begin{figure}[!htb]
    \centering
    \includegraphics[scale=1]{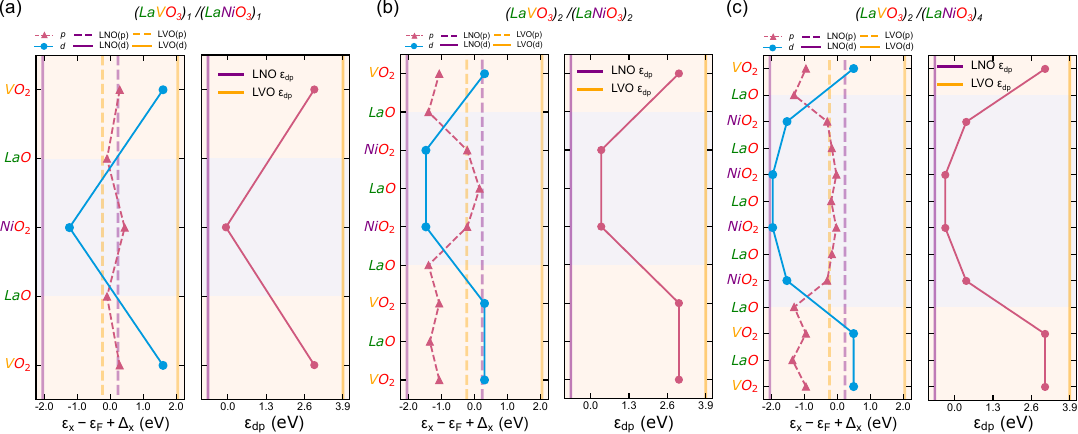}
    \caption{\textbf{Layer-resolved orbital energies and charge transfer energies.} The data are shown for the (a) (LaVO$_3$)$_1$/(LaNiO$_3$)$_1$, (b) (LaVO$_3$)$_2$/(LaNiO$_3$)$_2$, and (c) (LaVO$_3$)$_2$/(LaNiO$_3$)$_4$ superlattices. For each system, the left subpanel shows the orbital energies with respect to $\epsilon_x-\epsilon_F+\Delta_x$ ($x=p,d$) where $\epsilon_F$ is the Fermi energy and the $\Delta_x$ are arbitrary shifts introduced for centralize the curves around the zero ($\Delta_p=2.6$ eV and $\Delta_d=0.95$ eV). The right subpanel displays the energy difference $\epsilon_{dp}$ as a function of the TM site along the $y$ direction of the superlattice. Vertical lines in both subpanels denote the corresponding bulk reference values for LaVO$_3$ and LaNiO$_3$. }
    \label{fig:energy_charge_transfer}
\end{figure}

Beyond these electrostatic shifts, as detailed in Fig.~\ref{fig:energy_charge_transfer}, the interfacial charge transfer fundamentally alters the local metal-oxygen covalent bonding. As evidenced in left panels of Fig.~\ref{fig:energy_charge_transfer}(a),(b) and (c), the system exhibits a layer layer dependent band shift ($\epsilon_x-\epsilon_F+\Delta_x$) of both the oxygen $2p$ and transition metal d states relative to their bulk references. This phenomenon is best captured by the spatial evolution of the effective charge-transfer energy $\epsilon_{dp}$, defined here as the energy difference between the local d and p band centers. As shown in right panels of Fig.~\ref{fig:energy_charge_transfer}(a),(b) and (c), $\epsilon_{dp}$ exhibits a continuous transition across the superlattice. Drawing parallel to the interfacial charge accommodation described by Beck and Ederer \cite{beck2019charge,beck2023interfacial} in correlated oxide heterostructures, this continuous profile of $\epsilon_{dp}$ serves as a metric of the electronic reconstruction.

The layer resolved energy profiles for the thicker (LVO)$_2$/(LNO)$_4$ SL shows that the lattice accommodation of the new valence states is not strictly confined to the immediate interfacial plane. As observed in Fig.~\ref{fig:pdos_layer_complete}(c), the O $2p$ band center gradually recovers from $-2.87$ eV at the immediate NiO$_2$ interface to $-2.65$ in the bulk-like NiO$_2$ layers. This behavior indicates that the charge transfer effects are screened within a length scale of approximately 3 unit cells. In the other hand, the ultrathin (LVO)$_1$/(LNO)$_1$ SL are under a extreme confined regime, once every transition metal site is strictly localized at the interface, preventing to recovery the bulk-like state, stabilizing the emergent magnetic order.

\section{Structural evolution of TM octahedra and  charge transfer}

\begin{figure}[!htb]
    \centering
    \includegraphics[scale=1.0]{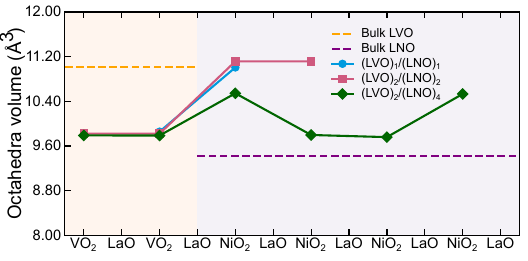}
    \caption{\textbf{Spatial evolution of local octahedral volumes.} The calculated volumes of the VO$_2$ and NiO$_2$ octahedra are shown across the transition metal oxides layers for the (LNO)$_1$, (LVO)$_2$/(LNO)$_2$ and (LVO)$_2$/(LNO)$_4$ superlattices. The horizontal dashed lines indicate the octahedra volumes for bulk LaVO$_3$ and LaNiO$_3$.}
    \label{fig:octahedral_volume}
\end{figure}

In Fig.~\ref{fig:octahedral_volume} we show the octahedra volume for the distinct SLs.

Compared to the bulk structures (indicated by horizontal dashed lines), a distinct volumetric expansion of the NiO$_6$ octahedra and a concomitant contraction of the VO$_6$ octahedra are observed at the interface regions. This volume variation provides a structural signature of the interfacial charge transfer. 

\section{Non-relativistic spin splitting in bulk LaVO$_3$}

We calculated the band structure and the correspondent spin splitting for the pristine bulk LaVO$_3$. The calculation was performed using the low temperature monoclinic structure ($P2_1/a$ space group) with converged $C$-AFM ground state.

\begin{figure}[H]
    \centering
    \includegraphics[scale=1]{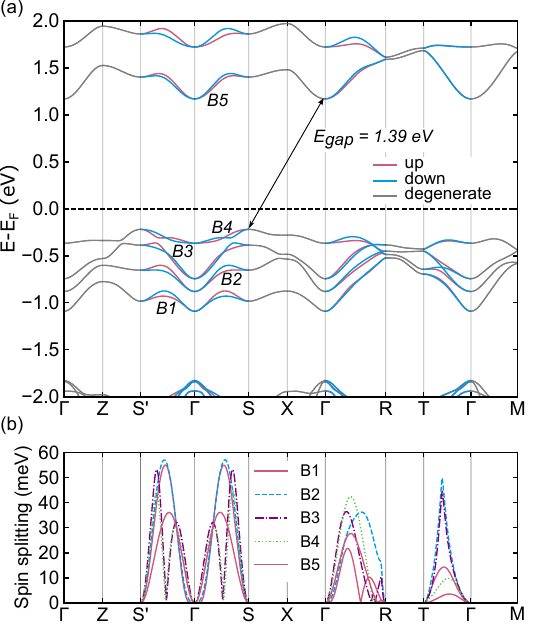}
    \caption{\textbf{Spin-resolved electronic band structure of bulk LaVO$_3$ (monoclinic $P2_1/a$ space group).} (a) Calculated DFT+U band structure for the $C$-type AFM ground state. The pin up and down channels are colored in red and blue, respectively. The system shows a band gap of $1.39$ eV. (b) Momentum-resolved magnitude of the non-relativistic spin splitting $\Delta E(\mathbf{k})$ for each denoted band ($B1$-$B5$).}
    \label{fig:bands_LVO_BULK}
\end{figure}

As shown in Fig.~\ref{fig:bands_LVO_BULK}(a), the bulk system is an insulator with a calculated band gap of $1.39$ eV. This value is in agreement with the optical gap measured experimentally of $1.1$ eV for the pristine LaVO$_3$ \cite{arima1993variation,assmann2013oxide}. The spin degeneracy is lifted along the $S'-\Gamma-S$, $\Gamma-R$ and $T-\Gamma$ directions, in agreement to the predicted altermagnetism nature of LaVO$_3$ on a $3d^2$ occupation, as recently reported in literature \cite{naka2025altermagnetic}. As quantified in Fig.~\ref{fig:bands_LVO_BULK}(b), the maximum spin splitting observed reaches approximately $57.1$ meV. By comparing this bulk reference to the SL studied in the main text reveals the importance of SL to tuning the materials properties. The interfacial electronic reconstruction and the octahedra distortions in the (LVO)$_1$/(LNO)$_1$ SL enhance the non-relativistic spin splitting, reaching to $117.9$ eV.

\section{Wannier bands}

\begin{figure}[htb]
    \centering
    \includegraphics[scale=1.2]{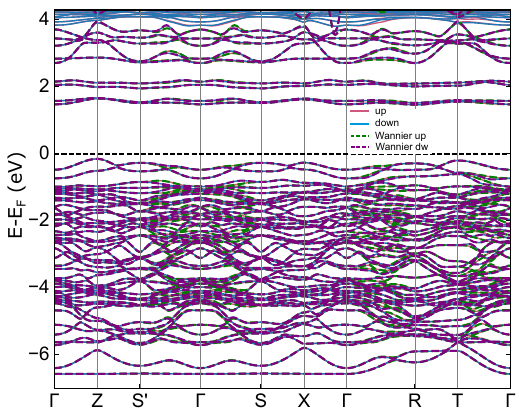}
    \caption{\textbf{Validation of the Wannier interpolated bands}. The Wannier interpolated bands (dashed lines) are compared with the first principles DFT+U bands (solid lines) calculated at $U_V=3.8$ eV and $U_Ni=4.8$ eV, showing an good representation over a $10.6$ eV span. The V-$d$, Ni-$d$ and O-$p$ orbitals were used in the wannier basis set.}
    \label{fig:wannier_bands}
\end{figure}

We cross-checked our band structure obtained using our $pd$ model Hamiltonian against the obtained DFT band structure. The comparison is shown in Fig.~\ref{fig:wannier_bands}.

\section{Symmetry constraints and the mixed $d$-wave spin-splitting form factor}

To understand the microscopic nature of the spin splitting, we analyze the band dispersion near the $\Gamma(\textbf{k}=0)$ point using a Taylor expansion constrained by the magnetic space group symmetries.

\subsection{(LVO)$_1$/(LNO)$_1$ superlattice}
In the (LVO)$_1$/(LNO)$_1$ superlattice, the system is centrosymmetric (space group $P2_1/c$). Because the collinear magnetic order preserves global spatial inversion ($\mathcal{P}$) within each spin channel independently, the energy dispersion must be an even function of the crystal momentum, $E_{\sigma}(\textbf{k})= E_{\sigma}(-\textbf{k})$. This forces all odd-order terms in the Taylor expansion to vanish. Expanding the energy to second order, we obtain:

\begin{equation}
    E_{\sigma}(\textbf{k})=c_{o,\sigma} + c_{xx,\sigma}k^2_x + c_{yy,\sigma}k^2_y + c_{zz,\sigma}k^2_z + c_{xy,\sigma}k_xk_y + c_{xz,\sigma}k_xk_z + c_{yz,\sigma}k_yk_z + \mathcal{O}
\end{equation}

The spin splitting is defined as the energy difference between opposite spins:

\begin{equation}
    \Delta E(\textbf{k})=E_{\uparrow}(\textbf{k}) - E_{\downarrow}(\textbf{k})
\end{equation}

Defining $\Delta c_{ij}=c_{ij,\uparrow}-c_{ij,\downarrow}$, the spin splitting takes the general form:

\begin{equation}
    \Delta E_{\sigma}(\textbf{k})=\Delta c_{o,\sigma} + \Delta c_{xx,\sigma}k^2_x + \Delta c_{yy,\sigma}k^2_y + \Delta c_{zz,\sigma}k^2_z + \Delta c_{xy,\sigma}k_xk_y + \Delta c_{xz,\sigma}k_xk_z + \Delta c_{yz,\sigma}k_yk_z + \mathcal{O}
\end{equation}

The altermagnetic spin splitting is governed by the nonsymmorphic operations that swap the opposite spin sublattices. Specifically, the $c$-glide plane ($M_z$) combined with time-reversal connects the spin up and spin down states. Applying this spin swapping operation ($k_z \rightarrow -k_z$ and $\Delta E\rightarrow -\Delta E$) imposes the symmetry constraint:

\begin{equation}
    \Delta E(\textbf{$k_x,k_y,k_z$})= - \Delta E(\textbf{$k_x,k_y,-k_z$})
\end{equation}

Inserting the Taylor expansion:

\begin{equation}
    \begin{split}
        &\Delta c_{o,\sigma} + \Delta c_{xx,\sigma}k^2_x + \Delta c_{yy,\sigma}k^2_y + \Delta c_{zz,\sigma}k^2_z + \Delta c_{xy,\sigma}k_xk_y + \Delta c_{xz,\sigma}k_xk_z + \Delta c_{yz,\sigma}k_yk_z =\\
        &=- \left[ \Delta c_{o,\sigma} + \Delta c_{xx,\sigma}k^2_x + \Delta c_{yy,\sigma}k^2_y + \Delta c_{zz,\sigma}(-k_z)^2 + \Delta c_{xy,\sigma}k_xk_y + \Delta c_{xz,\sigma}k_x(-k_z) + \Delta c_{yz,\sigma}k_y(-k_z) \right]
    \end{split}
\end{equation}

For this equality to hold for any arbitrary \textbf{k}, all coefficients of terms that are even with respect to $k_z$ must be identically zero ($\Delta c_{o,\sigma} = \Delta c_{xx,\sigma} = \Delta c_{yy,\sigma} =\Delta c_{zz,\sigma} = \Delta c_{xy,\sigma}=0$). Thus, the splitting takes the form:

\begin{equation}
    \Delta E(\textbf{k})= \Delta c_{xz}k_xk_z + \Delta c_{yz}k_yk_z
\end{equation}

Similarly, the two-fold screw axis ($C_{2z}$) acts as a second spin-swapping operation ($k_x \rightarrow -k_x$, $k_y \rightarrow -k_y$ and $\Delta E\rightarrow -\Delta E$). This imposes the constraint:

\begin{equation}
    \Delta E(\textbf{$k_x,k_y,k_z$})= - \Delta E(\textbf{$-k_x,-k_y,k_z$})
\end{equation}

This identity is automatically satisfied, confirming that both $k_xk_z$ and $k_yk_z$ terms are allowed by the full space group. Therefore, up to second order, the momentum-dependent spin splitting follows:

\begin{equation}
    \Delta E(\textbf{k})= \alpha k_xk_z + \beta k_yk_z
    \label{eq:analytical_1_1}
\end{equation}

where $\alpha \equiv \Delta c_{xz}$ and $\beta \equiv \Delta c_{yz}$. This dependence shows that the symmetry monoclinic environment allows a mixed $d$-wave form factor rather than a pure single one.

\begin{figure}[!htb]
    \centering
    \includegraphics[scale=1.2]{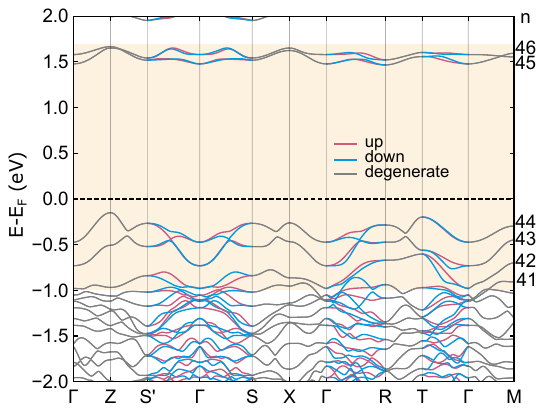}
    \caption{\textbf{Band structure of the (LVO)$_1$/(LNO)$_1$ superlattice.} The highest occupied valence bands and the lowest conduction bands, indexed from 41 to 46, are highlighted.}
    \label{fig:interval_fitting}
\end{figure}

To confirm this $d$-wave mixed symmetry form factor, we fitted the splittings of the highest occupied valence bands (band indices 41 to 46), as illustrated in Fig.~\ref{fig:interval_fitting}, to the analytical expression in Eq.~\ref{eq:analytical_1_1}. These bands correspond to the strongly hybridized V-$t_{2g}$, Ni-$e_g$, and O-$2p$ states that host the altermagnetic phase. We constrained our numerical fitting to a specific momentum interval around $\Gamma$ where the second-order Taylor expansion remains a good approximation. The extracted altermagnetic constants ($\alpha$ and $\beta$) for each band and their respective coefficients of determination ($R^2$) are summarized in Table~\ref{tab:coefficients}.

\begin{table}[!htb]
    \centering
    \caption{\textbf{Altermagnetic d-wave form factor constants.} Calculated coefficients ($\alpha$ and $\beta$) for the mixed $d_{xz} + d_{yz}$ form factor model in the (LVO)$_1$/(LNO)$_1$ superlattice evaluated near the $\Gamma$ point. The standard errors and the coefficient of determination ($R^2$) for each analyzed band are included.}
    \renewcommand{\arraystretch}{1.2}
    \setlength{\tabcolsep}{8pt}
    \begin{tabular}{@{}cccc@{}}
        \toprule \hline 
        Band Index & $\alpha$ & $\beta$ & $R^2$ \\ 
        \midrule
         41 & $(10.10\pm0.01)$ &  $(3.65\pm0.01)$ & 0.9549 \\
        42 & $(-2.554\pm0.006)$ & $(-1.094\pm0.006)$ & 0.9919 \\
        43 & $(-0.7497\pm0.0003)$ & $(3.5482\pm0.0003)$ & 0.9996 \\
        44 & $(1.2723\pm0.0004)$ & $(-0.5853\pm0.0004)$ & 0.9959 \\
        45 & $(0.73934\pm0.00004)$ & $(-0.67547\pm0.00004)$& 0.9996 \\
        46 & $(-1.48713\pm0.00005)$ & $(0.39585\pm0.00005)$ & 0.9997 \\
        \bottomrule 
    \end{tabular}
    \label{tab:coefficients}
\end{table}

\bibliography{main}